\documentclass[twocolumn]{elsarticle}

\usepackage{graphicx}
\graphicspath{{figures/}} 
\usepackage[utf8]{inputenc}
\usepackage{amssymb}
\usepackage{amsmath}
\usepackage{lineno}

\usepackage{hyperref}
 \hypersetup{
     colorlinks=true,
     linkcolor=blue,
     filecolor=blue,
     citecolor = black,      
     urlcolor=cyan,
     }

\newcounter{bla}

\journal{Computer Physics Communications}

\newcommand{\udkm}{\textsc{udkm1Dsim}~}
\newcommand{\code}[1]{\mintinline{python}{#1}}
\newcommand{\figurewidth}{0.4}

\usepackage[frozencache,cachedir=./minted]{minted}
\setminted{fontsize=\footnotesize}

\begin{document}

\onecolumn

\begin{frontmatter}

\title{\udkm - A Python toolbox for simulating 1D ultrafast dynamics in condensed matter}

\author{Daniel Schick\corref{author}}

\address{Max-Born-Institut für Nichtlineare Optik und Kurzzeitspektroskopie, Max-Born-Straße 2A, 12489 Berlin, Germany}



\cortext[author]{Corresponding author. \textit{E-mail address:} schick@mbi-berlin.de}

\begin{abstract}
The \udkm toolbox is a collection of Python classes and routines to simulate the thermal, structural, and magnetic dynamics after laser excitation as well as the according X-ray scattering response in one-dimensional sample structures.
The toolbox provides the capabilities to define arbitrary layered structures on the atomic level including a rich database of element-specific physical properties.
The excitation of dynamics is represented by an $N$-temperature-model which is commonly applied in ultrafast physics.
Structural dynamics due to thermal stresses are calculated by a linear-chain model of masses and springs.
The implementation of specific magnetic dynamics can be easily accomplished by the user employing a generalized magnetization interface class.
The resulting X-ray diffraction response is computed by kinematical or dynamical X-ray theory which can also include polarization-dependent magnetic scattering.
The \udkm toolbox is highly modular and allows for injecting user-defined inputs at any step within the simulation procedure.
\end{abstract}

\newpage

\begin{keyword}
ultrafast dynamics \sep
multilayer absorption \sep
heat diffusion \sep
$N$-temperature model \sep
coherent phonons \sep
magnetization dynamics \sep
thermoelasticity \sep
magnetoelasticity \sep
kinematical X-ray scattering \sep
dynamical X-ray scattering \sep
dynamical magnetic X-ray scattering

\end{keyword}

\end{frontmatter}


\newpage

{\bf PROGRAM SUMMARY}

\begin{small}
\noindent
{\em Program Title:}
udkm1Dsim \\
{\em CPC Library link to program files:}
(to be added by Technical Editor) \\
{\em Developer's repository link:}
https://github.com/dschick/udkm1Dsim \\
{\em Code Ocean capsule:} (to be added by Technical Editor)
\\
{\em Licensing provisions(please choose one):}
MIT \\
{\em Programming language:}
Python \\
{\em Journal reference of previous version:} 
Computer Physics Communications Volume 185, Issue 2, February 2014, Pages 651-660 \\
{\em Does the new version supersede the previous version?:}
yes \\
{\em Reasons for the new version:}
The toolbox has been ported from {\sc MATLAB} (MathWorks Inc.) to Python and is based exclusively on free and open-source components.
Moreover, new features have been added that allow for a broader applicability of the toolbox.\\
{\em Summary of revisions:}\\
Porting to Python.\\
Introduction of amorphous layers in the sample structures.\\
Add magnetization property to atoms and layers.\\
Multilayer formalism to calculate laser absorption.\\
New magnetization class to allow for user-defined magnetization dynamics.\\
New resonant magnetic X-ray scattering employing dynamical X-ray theory.\\
Calculation of X-ray scattering as function of photon energy and scattering vector.\\
{\em Nature of problem(approx. 50-250 words):} Simulate the thermal, structural, and magnetic dynamics of 1D sample structures due to an ultrafast laser excitation and compute the corresponding transient (magnetic) X-ray scattering response.\\
{\em Solution method(approx. 50-250 words):} The program provides an object-oriented toolbox for building arbitrary layered 1D crystalline/amorphous sample structures including a rich database of element-specific parameters.
The excitation, thermal transport, and lattice dynamics are simulated utilizing SciPy’s ODE solver.
Magnetization dynamics can be introduced by the user employing a magnetization interface class.
The dynamical (magnetic) X-ray scattering is computed by a matrix formalism that can be parallelized.\\
{\em Additional comments including restrictions and unusual features (approx. 50-250 words):} The program is restricted to 1D sample structures.
Phonon dynamics only include longitudinal acoustic phonons (sound waves).
Magnetization dynamics have to be defined by the user.
X-ray scattering only allows for symmetrical and co-planar geometries due to the 1D nature of the toolbox.
The program is highly modular and allows the inclusion of user-defined inputs at any time of the simulation procedure.\\

\end{small}

\newpage

\twocolumn

\section{Introduction}

The investigation of electronic, magnetic, and structural dynamics in solid state physics has made great progress during the last decades due to the increasing availability of ultrashort electron and light pulses in a broad spectral range from THz to hard X-rays at large-scale facilities as well as in the lab.
One of the major goals of these experiments is to follow the coupling of different degrees of freedom on the relevant time and length scales.
Recent examples reveal the complexity of these experiments even for following only two sub-systems, such as the coupling of electron and lattice dynamics \cite{Waldecker2016, schi2014a, nico2011a}, of electron and spin dynamics \cite{Willems2020, You, stam2007a}, or the coupling of spin and lattice dynamics \cite{Jal2017, Henighan2016, Reppert2016}.

In order to understand and interpret such experimental data, one relies on a bunch of simulations for modeling and fitting, which are available as software tool-kits or as published formalisms.
A list of required dynamic simulations might include the simplistic $N$-temperature model (NTM) as initially proposed by \citet{anis1974a} as well as its modifications and implementations \cite{Waldecker2016, Koopmans2010, Alber2020}, coherent phonon dynamics in a masses-and-spring model \cite{thom1986a, herz2012b}, molecular dynamics for simulations of spin lattice coupling \cite{spilady, MA2016350}, magnetic simulations combining a 2-temperature-model with the Landau-Lifschitz-Bloch equation \cite{Atxitia2017}, or full magnetic simulations \cite{Ubermag, mumax3, oommf, Vansteenkiste2014}.
Other aspects can be the spatial absorption profile of the laser excitation in the sample \cite{Ohta1990, LeGuyader2013} and, moreover, the calculation of the actual observable, e.g. the scattered light intensity in the framework of kinematical or dynamical scattering theories \cite{Kriegner2013, step1998a, Windt1998, alsn2001a, Elzo2012, Macke2014} with possible inclusion of resonant and magnetic scattering effects.
The implementation of the above mentioned formalisms or the usage and adaption of available external software packages is very time-consuming and each piece of software covers only a very limited aspect of real time-resolved experiments.
To that end, the need for a generic, modular, and open-source toolbox that allows for combining many of the above mentioned functionalities is obvious.

The \udkm toolbox allows to create arbitrary one-dimensional (1D) structures made of crystalline and/or amorphous layers, including stochiometric mixtures, typically on the nanometer length-scale.
These 1D structures hold all relevant material information such as structural, elastic, thermal, magnetic, and optical parameters.
The toolbox allows for calculating thermal, structural, and magnetic dynamics on these 1D structures utilizing an NTM, a linear masses-and-springs-model, as well as an interface for user-defined magnetization dynamics, respectively.
Finally, different types of light-scattering theories can be applied to retrieve the static as well as transient response from these sample structures due to the above mentioned dynamics, similar to real pump-probe experiments.
With that the generally non-linear dependence of the actual observable (scattered light intensity) and the physical quantity of interest (temperature, strain, magnetization, ...) can be revealed.
This includes also methods to apply realistic instrumental broadening to the simulated results for better comparison with experimental data.
For now the \udkm toolbox focus on light scattering from the extreme ultra-violet (XUV) up to hard X-rays but can be easily extended to larger wavelength radiation or totally different probing techniques that can be modeled in 1D.

The original \textsc{matlab} (MathWorks Inc.) \udkm toolbox \cite{schi2013c} has been successfully used in 40 publications (Feb. 2021) and is still available for download \cite{udkmML}.
However, due to Python's increasing performance, popularity, and availability the new version of the \udkm toolbox has been ported to this programming language.
At the same time, the project has moved to \href{https://github.com/dschick/udkm1Dsim}{\nolinkurl{github.com/dschick/udkm1Dsim}} to provide full version control, issue, and feature tracking, as well as project management capabilities in order to allow for better collaboration between users and developers.
This also includes automatic code validation and unit testing, as well as source-code-based generation of the documentation at \href{https://udkm1Dsim.readthedocs.org}{\nolinkurl{udkm1Dsim.readthedocs.org}} as part of the continuous integration (CI) concept.

Despite of the major step of porting the \udkm toolbox to Python there are several new features included in this release while only a few minor features of the old version have been dropped.
Accordingly, existing simulations in \textsc{matlab} (MathWorks Inc.) are easily portable to this new Python version as the general concept and syntax of the toolbox has been unaffected.
The new functionalities of the release include: the introduction of amorphous layers in addition to crystalline unit cells; magnetic properties of atoms including magnetic scattering factors; a multilayer formalism for calculating laser absorption profiles \cite{Ohta1990, LeGuyader2013}; a new magnetization dynamics module; a dynamical magnetic scattering formalism \cite{Elzo2012}; as well as a unified interface to calculate light scattering as function of scattering vector and photon energy.

Most of the underlying physics has already been described in the initial version of the \udkm toolbox \cite{schi2013c} as well as in the \href{https://udkm1dsim.readthedocs.io/en/latest/api.html}{API documentation} of the corresponding modules.
Therefore we will concentrate mainly on the new features/physics as well as on the slight changes of the workflow in Python.
After a general description of the implementation and workflow of the \udkm toolbox, we will introduce the new release features within an exemplary simulation of laser-induced dynamics in a magnetically-coupled superlattice of Fe and Cr layers.

As a convention throughout this document all code is written in typewriter font (\code{code = [1, 2, 3]}). 

\section{Implementation \& Workflow}

\begin{figure}[tb]
    \centering
    \includegraphics[width=\figurewidth\textwidth]{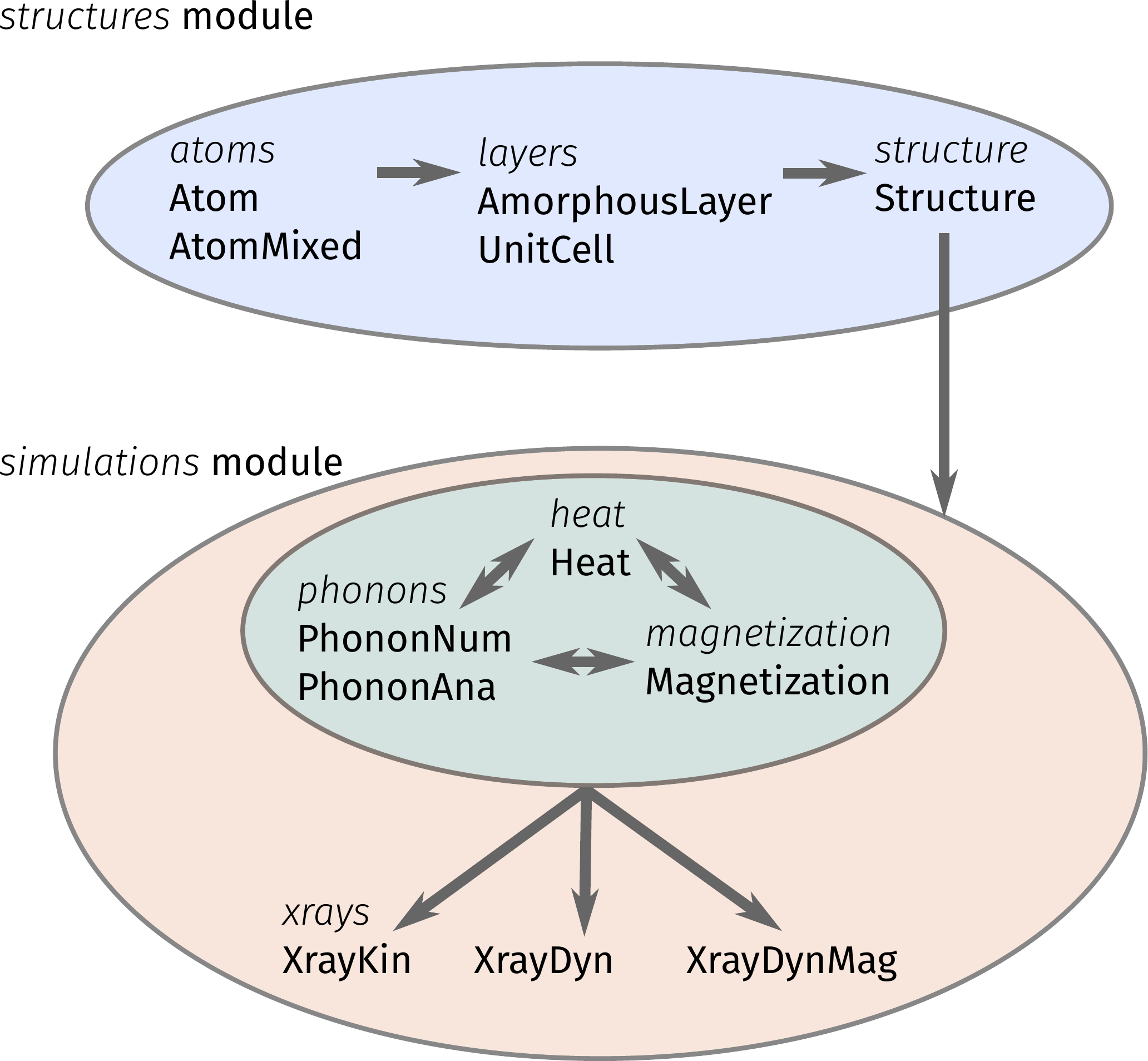}
    \caption{Workflow of the \udkm toolbox.
    (Sub-)Module names are \textit{italic} and class names are in \textbf{bold} letters.
    All classes of the \code{simulations} module require a \code{Structure} object on initialization.
    }
    \label{fig:scheme}
\end{figure}

The structure of the \udkm toolbox as a Python module tries to reflect the physical reality of the modelled experiments, see Fig.~\ref{fig:scheme}.
In the beginning a \code{Structure} object needs to be build out of any number and combination of \code{AmorphousLayer} and/or \code{UnitCell} objects which themselves need to consist of one or more \code{Atom}/\code{AtomMixed} objects.
The \code{Structure} holds all relevant physical parameters which are required for the actual static and dynamic simulations.
Within the \code{simulations} module different types of dynamic simulations can be carried out on the \code{Structure} object, e.g. \code{Heat, PhononNum/Ana, Magnetization} simulations.

In general, the \code{Heat} class models the laser-excitation and the resulting energy-flow by an NTM.
The resulting phonon and magnetization dynamics can be calculated sequentially or in any user-defined class, that allows for coupling of the different degrees of freedom.
The modular structure of the in- and outputs of the simulation classes allow for easy exchange or alteration of the simulation results with user-defined data or external calculations.
The classes of the \code{xray} module allow for calculation of static symmetric X-ray scattering from the given sample \code{Structure} as function of photon energy, scattering vector, as well as of light polarization.
In addition, the results from the dynamic \code{Heat, PhononNum/Ana, Magnetization} simulations can be input to the X-ray scattering calculation in order to retrieve the dynamic scattering response of the excited sample.
This functionality of the \udkm toolbox is of major importance, since the transient change of the actual physical quantity of interest, such as strain or magnetization, must not be proportional to the actual observable - the scattered light intensity.

\begin{figure}[tb]
    \centering
    \includegraphics[width=\figurewidth\textwidth]{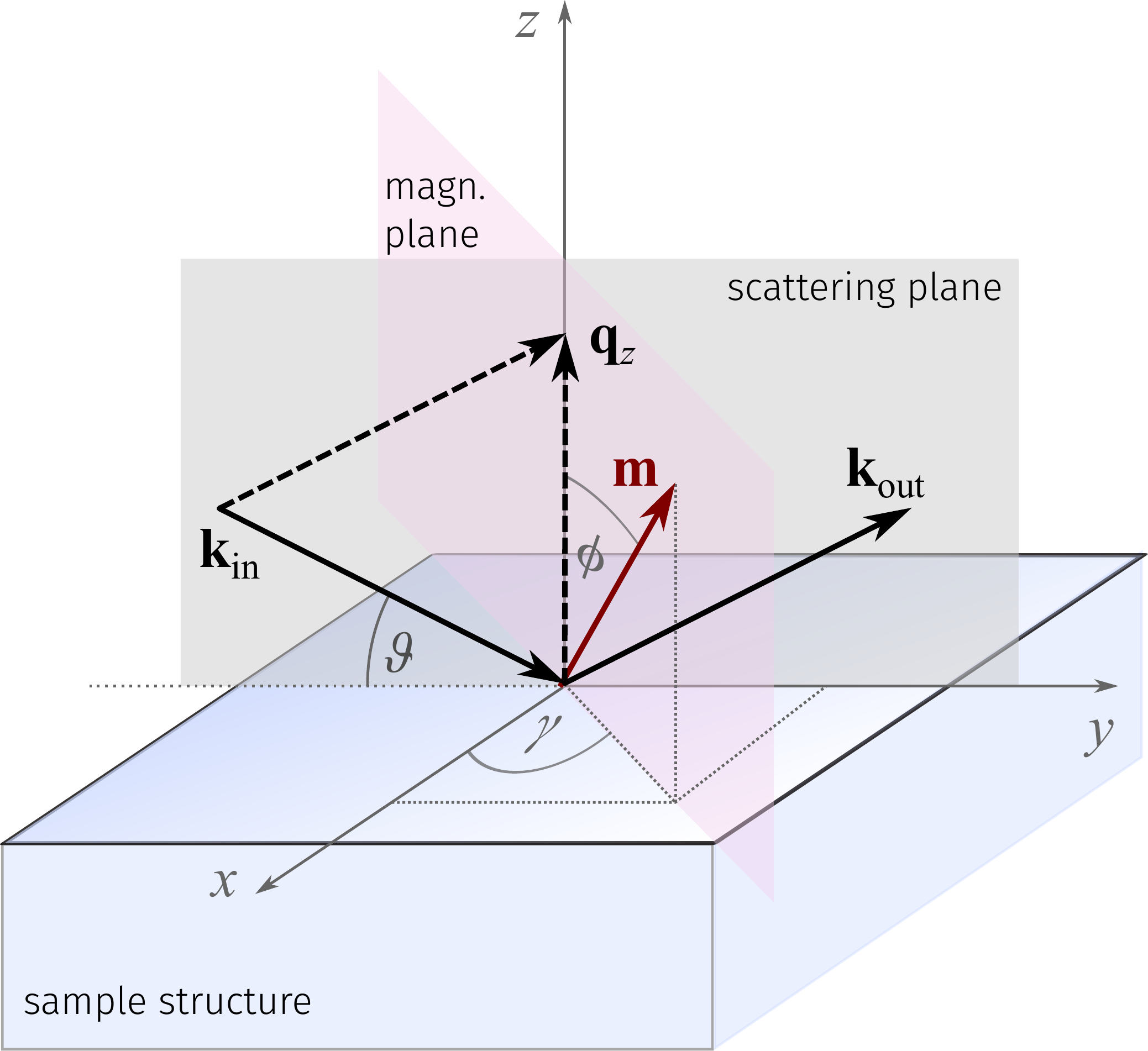}
    \caption{Experimental scheme of the \udkm toolbox.
    The co-planar scattering plane contains the scattering vector $\vec{q_z} = \vec{k}_\text{out} -\vec{k}_\text{in}$.
    The magnetization $\vec{m}$ of the individual \code{Atom} objects can be defined in 3D by an amplitude and the according angles $\gamma$ and $\phi$.
    The out-of-plane coordinate $z$ points away from the sample surface. 
    }
    \label{fig:experiment}
\end{figure}

The common experimental scheme is sketched in Fig.~\ref{fig:experiment} and shows the general definition of coordinates and angles that are used throughout the \udkm toolbox.
Due to the limitation to 1D sample structures the scattering plane must be co-planar to the sample surface and only symmetrical scattering with a scattering vector $\vec{q_z} = \vec{k}_\text{out} -\vec{k}_\text{in}$ can be calculated in the X-ray simulations.
Here the norm of the scattering vector $q_z = \frac{4\pi}{\lambda} \sin{\vartheta}$ is defined by the X-ray wavelength $\lambda$ and the incidence angle $\vartheta$.
The multilayer absorption uses the same definition as shown in  Fig.~\ref{fig:experiment} for the laser incidence angle.
The magnetization $\vec{m}$ of each \code{Atom} is defined by the according angles $\gamma$ and $\phi$.
The magnetization can point into all directions in 3D in order to allow for any overlap with the incident X-ray polarization.
Since the sample structure is defined from top to bottom the distances within the sample start at its surface and proceed along $-z$ direction.

All classes of the \code{simulations} module share an advanced caching mechanism which allows to save computationally heavy results to a user-defined path for later reuse.
If enabled, the individual simulations generate a unique hash for only the relevant physical parameters that affect the requested calculation.
The result of the calculation will then be saved and tagged with this unique hash.
It will be automatically reloaded if the simulation is carried out again with the exact same parameter set.

The \udkm toolbox fully relies on SI units within its internal calculations.
However, the external user interface allows to define all relevant parameters as physical quantities using the \code{Pint} package \cite{Pint}.
Accordingly, the unit conversion to proper SI units will be done automatically within the code.
If no units are defined, parameters are assumed to be already in SI units.

\section{Example}

We choose to simulate the laser-driven structural and magnetic dynamics in an antiferromagnetically (AFM) coupled Fe/Cr superlattice (SL) \cite{Grunberg1986} as an example to demonstrate the recent capabilities of the \udkm toolbox.
The test sample consists of 20x(Fe+Cr) on a Si substrate with a layer thickness of $d_\text{Fe} = d_\text{Cr} = 1$~nm.
In the Fe/Cr SL the individual Fe layers are ferromagnetically aligned in the plane of the sample.
In dependence of the Cr layer thickness the adjacent Fe layers can align anti-parallel to each other \cite{Fullerton1993}.
Generally so-called SL Bragg peaks will appear in the X-ray diffraction due to the artificial periodicity of an Fe/Cr doublelayer (DL) \cite{holy1999a}.
For the special case of resonant magnetic scattering, e.g. at the Fe $L_2$ and $L_3$ edges, the doubled AFM periodicity of two DLs is observable and leads to pure magnetic Bragg peaks at half order of the structural Bragg peaks \cite{Nefedov2005}.

The AFM Fe/Cr SL is already a very complex example and the \udkm toolbox will be far from covering all involved dynamics here.
However, it can provide important insights into the laser excitation using the multilayer absorption formalism, the energy transport on ultrafast timescales using a heat diffusion model, resulting coherent acoustic phonons due to thermal stresses, and a simplified demagnetization model due the laser-driven temperature increase.
Moreover, the resulting magnetic X-ray scattering response allows for a direct comparison between experiment and simulations.

After successful installation of the \udkm toolbox, it can be easily imported in any Python script or Jupyter notebook:
\begin{minted}{python}
import udkm1Dsim as ud
u = ud.u
\end{minted}
Here \code{u} is the so-called unit-registry based on the \code{Pint} package \cite{Pint} which is used internally in the toolbox and must be used externally to allow for physical quantities and automatic unit conversion.

\subsection{Structure}

In the first step the \code{Atom} and \code{AmorphousLayer} objects are created to build the sample \code{Structure}:
\begin{minted}{python}
Fe_right = ud.Atom('Fe', mag_amplitude=1,
    mag_phi=90*u.deg, mag_gamma=90*u.deg,
    atomic_form_factor_path='./Fe.cf')
Fe_left = ud.Atom('Fe', mag_amplitude=1,
    mag_phi=90*u.deg, mag_gamma=270*u.deg,
    atomic_form_factor_path='./Fe.cf')
Cr = ud.Atom('Cr')
Si = ud.Atom('Si')
\end{minted}
The two atoms \code{Fe_left} and \code{Fe_right} represent the two different anti-parallel magnetization states in the sample.
The parameters \code{mag_amplitude}, \code{mag_phi}, and \code{mag_gamma} define the magnitude and pointing of the Fe magnetization in accordance to Fig.~\ref{fig:experiment}.
Here the magnetization can be understood as the so-called macro-spin as within the Landau-Lifschitz-Bloch theory \cite{Atxitia2017}, that allows for a temperature-dependent variation of the magnetization amplitude.
The element-specific atomic scattering factors are taken from the Chantler \cite{Chantler2001} or Henke \cite{henk1993a} tables which are both included in the toolbox.
The magnetic form factors are taken from the Dyna project \cite{dyna}.
In case an \code{atomic_form_factor_path} or \code{magnetic_form_factor_path} to a user-defined data file is given, the according values will be used instead.

\begin{figure*}[tb]
    \centering
    \includegraphics[width=0.99\textwidth]{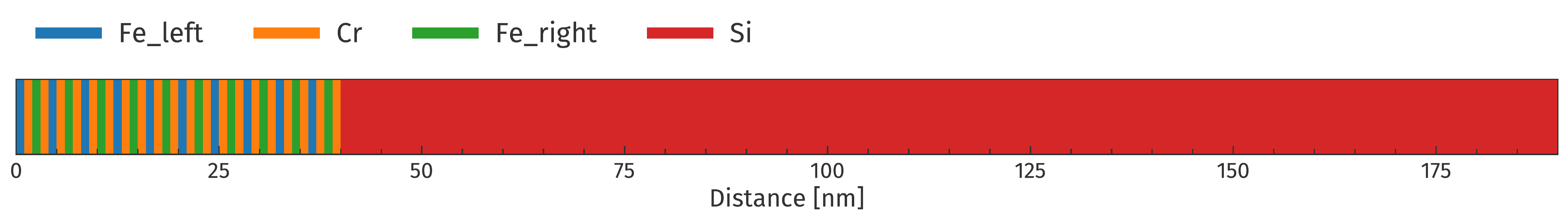}
    \caption{Automatic visualization of the AFM-coupled Fe/Cr SL \code{Structure}.
    The blue and green regions indicate \emph{left} and \emph{right} magnetized Fe layers, respectively.
    The orange and red regions indicate Cr and Si layers, respectively.
    The distances within the \code{Structure} are along the $-z$ coordinate.
    }
    \label{fig:structure_visualize}
\end{figure*}

\code{Layer} objects of Fe, Cr, and Si as substrate material are initialized as amorphous layers and by providing a dictionary of the relevant physical parameters:
\begin{minted}{python}
prop_Fe = {}
prop_Fe['heat_capacity'] = 449*u.J/u.kg/u.K
prop_Fe['therm_cond'] = 80*u.W/(u.m *u.K)
prop_Fe['lin_therm_exp'] = 11.8e-6
prop_Fe['sound_vel'] = 4.910*u.nm/u.ps
prop_Fe['opt_ref_index'] = 2.9174+3.3545j

layer_Fe_left = ud.AmorphousLayer('Fe_left',
    'Fe left amorphous', thickness=0.2*u.nm,
    density=7874*u.kg/u.m**3, atom=Fe_left, **prop_Fe)
layer_Fe_right = ud.AmorphousLayer('Fe_right',
    'Fe right amorphous', thickness=0.2*u.nm,
    density=7874*u.kg/u.m**3, atom=Fe_right, **prop_Fe)
...
layer_Cr = ud.AmorphousLayer('Cr', "Cr amorphous",
    thickness=0.2*u.nm, density=7140*u.kg/u.m**3,
    atom=Cr, **prop_Cr)
layer_Si = ud.AmorphousLayer('Si', "Si amorphous",
    thickness=0.2*u.nm, density=2336*u.kg/u.m**3,
    atom=Si, **prop_Si)
\end{minted}
The thickness of the basic Fe and Cr layers is chosen to be only 0.2~nm in order to allow for a finer spatial grid for the ordinary differential equation (ODE) solvers in the dynamics simulations.
The sample \code{Structure} is created of a nested double-layer \code{Structure} which allows to simplify rather complex sample geometries:
\begin{minted}{python}
S = ud.Structure('Fe/Cr AFM Super Lattice')
# create a sub-structure
DL = ud.Structure('Two Fe/Cr Double Layers')
DL.add_sub_structure(layer_Fe_left, 5) # = 1nm
DL.add_sub_structure(layer_Cr, 5) # = 1nm
DL.add_sub_structure(layer_Fe_right, 5) # = 1nm
DL.add_sub_structure(layer_Cr, 5) # = 1nm
# add substructure to structure
S.add_sub_structure(DL, 10)
S.add_sub_structure(layer_Si, 500) # = 100nm
\end{minted}
The sample \code{Structure} can be easily visualized by the command \code{S.visualize()}, see Fig.~\ref{fig:structure_visualize}

\subsection{Simulations}

All transient simulations of the \udkm toolbox need a spatial and temporal grid defined in the beginning:
\begin{minted}{python}
delays = numpy.r_[-1:20:0.1]*u.ps
_, _, distances = S.get_distances_of_layers()
\end{minted}
Here one can freely choose the temporal grid while the spatial grid is generally given by the sample geometry which can be easily accessed from the \code{Structure} object itself.

\subsubsection{Heat}
\begin{figure}[tb]
    \centering
    \includegraphics[width=\figurewidth\textwidth]{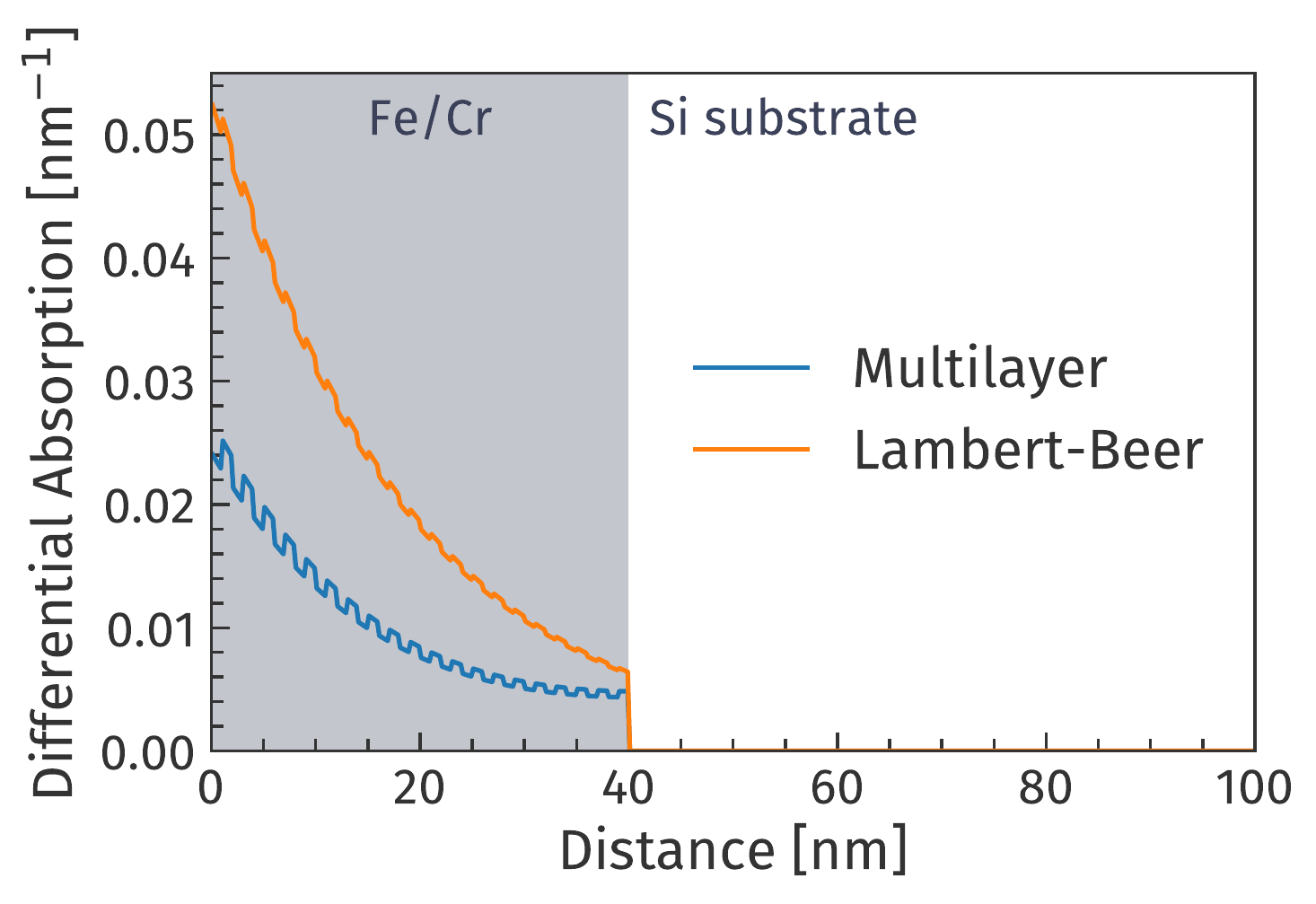}
    \caption{Differential absorption calculation comparing the Lambert-Beer law vs. the multilayer formalism.
    Despite internal multi-reflections, the multilayer formalism also includes reflection from the front surface of the sample.
    The absorption contrast between Fe and Cr layers is significantly different for both methods.
    }
    \label{fig:diff_abs}
\end{figure}

In order to simulate the laser-driven dynamics, we need to initialize a \code{Heat} object first and set all relevant excitation parameters.
\begin{minted}{python}
h = ud.Heat(S, True)
h.excitation = {'fluence': [20]*u.mJ/u.cm**2,
                'delay_pump':  [0]*u.ps,
                'pulse_width':  [0.1]*u.ps,
                'multilayer_absorption': True,
                'wavelength': 800*u.nm,
                'theta': 45*u.deg}
h.heat_diffusion = True
\end{minted}
The fluence $F_\text{incident}$ has to be input for normal incidence .
For the projection of the laser onto the sample surface, the angle of incidence $\vartheta$ is automatically taken into account, which is for $\vartheta = 45^\circ$ about 70~\% of $F_\text{incident}$.
In order to calculate the laser absorption profile, the user can either enable a multilayer absorption formalism \cite{Ohta1990, LeGuyader2013} or use the Lambert-Beer formula as fall-back.
Depending on the complexity of the sample geometry and the variation of the complex refractive index, the difference in the differential absorption d$A$/d$z$ between both algorithms can be significant, see Fig.~\ref{fig:diff_abs}.
The Lambert-Beer formula also neglects any reflection from the sample surface, while the multilayer formalism includes even multiple reflections inside the sample.
For the current sample the optical reflectivity is about 48~\% and the transmission is about 10~\%.
From these parameters one can calculate that the absorbed fluence in the sample is only about  $F_\text{abs}/F_\text{incident} = 29$~\%.

\begin{figure}[tb]
    \centering
    \includegraphics[width=\figurewidth\textwidth]{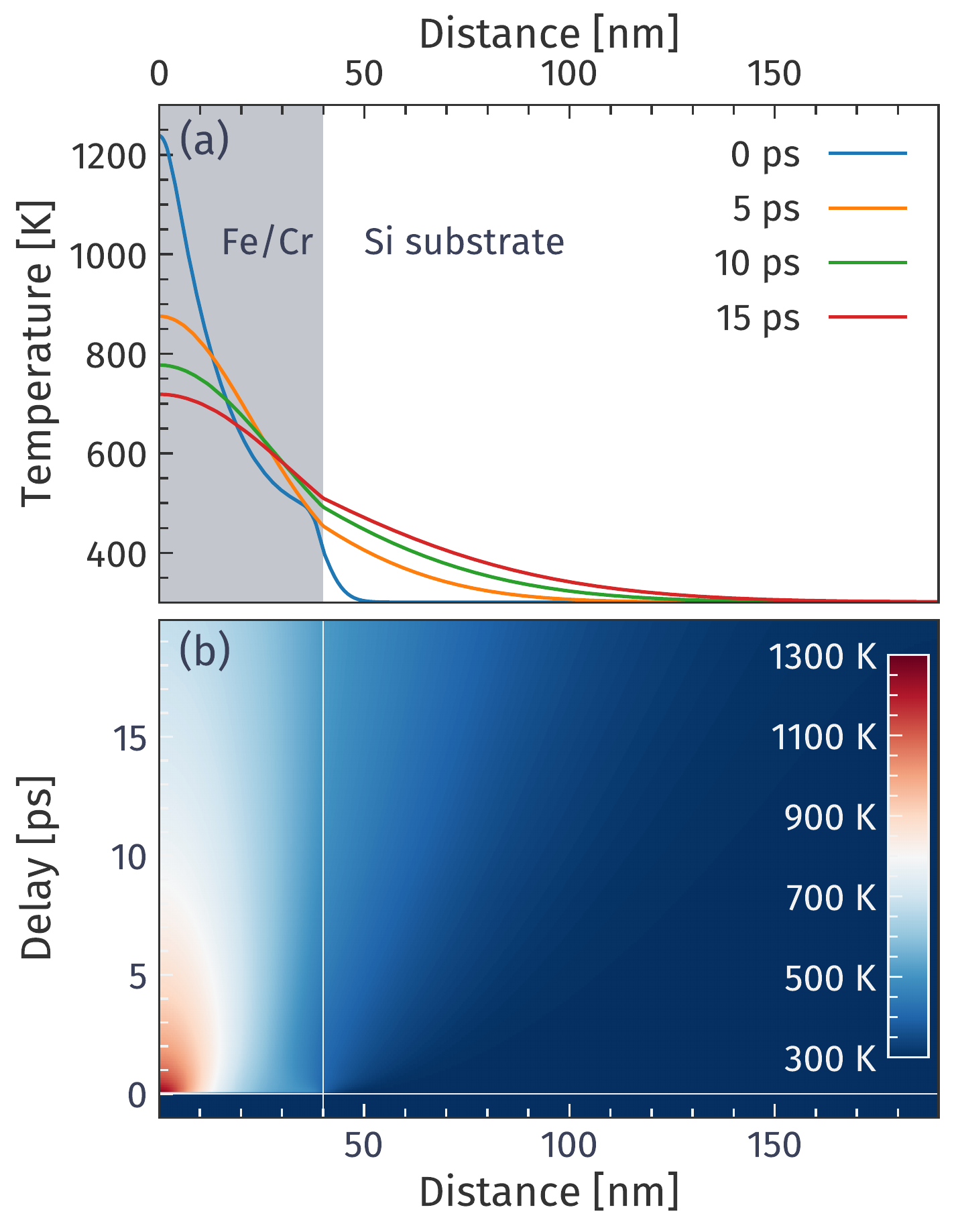}
    \caption{Spatio-temporal temperature map after laser excitation.
    The line-outs in (a) and the full color-coded map in panel (b) represent the relaxation of the laser-induced temperature gradient between absorptive Fe and Cr layers and the transparent Si substrate.
    Due to the finite laser pulse width of 100~fs the small temperature gradients between Fe and Cr layers within the first 40~nm are immediately relaxed.
    }
    \label{fig:heat}
\end{figure}

The calculation of the 1D heat diffusion equation on the above defined sample structure can then be simply called by:
\begin{minted}{python}
temp_map, delta_temp_map = h.get_temp_map(delays,
                                          300*u.K)
\end{minted}
Here the second argument of the \code{get_temp_map()} method is the homogeneous initial temperature of the sample which could be also an array containing initial temperatures for all individual layers.
The resulting spatio-temporal temperature map of the sample is shown in Fig.~\ref{fig:heat}.
In the background the \udkm toolbox will also take care of sufficient spatial and temporal interpolation at layer interfaces and pump events in order to improve the performance of the ODE solver dramatically.
This includes also the separation of temporal regions, that do not require the calculation of heat diffusion because full equilibrium conditions (no temperature gradients, no excitation, isolating boundary conditions).
More advanced settings like boundary conditions, multipulse excitations, or ODE settings can be found in the according examples of the \udkm toolbox.

\subsubsection{Numerical Phonons}

\begin{figure}[tb]
    \centering
    \includegraphics[width=\figurewidth\textwidth]{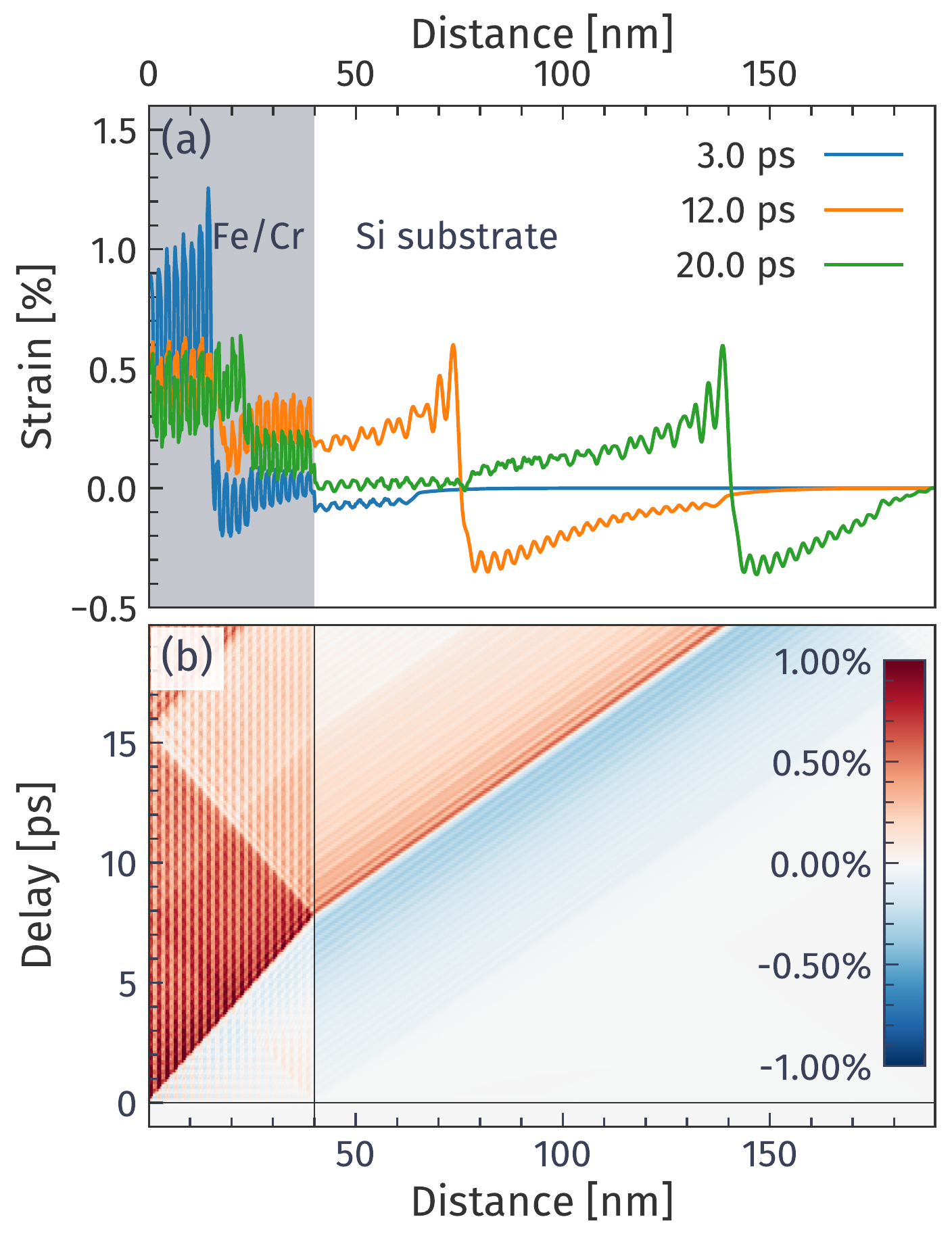}
    \caption{Spatio-temporal strain map due to the laser-excitation of coherent acoustic phonons.
    The full map in (b) features a complex pattern of SL phonons mode within the first 40~nm of the sample with tensile strain amplitudes of more than 1.0~\%.
    Panel (a) shows the bi-polar strain wave that is launched from the SL into the substrate at later delays.
    The bi-polar strain wave features additional oscillations that stem from the SL geometry.
    }
    \label{fig:phonons}
\end{figure}

Using the above calculated \code{temp_map} and its temporal derivative \code{delta_temp_map} it is easily possible to calculate the coherent acoustic phonon dynamics within the same sample \code{Structure} by using the numerical solver for the masses-and-spring model implemented in the \code{PhononNum} class:
\begin{minted}{python}
p = ud.PhononNum(S, True)
strain_map = p.get_strain_map(delays,
                              temp_map,
                              delta_temp_map)
\end{minted}
The resulting \code{strain_map} represents the strain for every layer at every delay point of the \code{Heat} simulation, see Fig.~\ref{fig:phonons}.

The map features complex oscillatory structures due to the excitation of so-called SL optical phonon modes or zone-folded acoustical phonons (ZFLAPs) \cite{herz2012b} within the Fe/Cr region.
These spatial and temporal oscillations of the strain are also imprinted on the bi-polar strain wave that is launched into the Si substrate at later delays.
One can easily read the sound velocity of Si from the slope of the color gradient in Fig.~\ref{fig:phonons}~(b) as about 8~nm/ps.
More details on the ODE settings, phonon damping, and non-linear phonon propagation can be found in the according examples of the \udkm toolbox. 

\subsubsection{Magnetization}

The simulation of magnetization dynamics is a rather complex topic on its own and the \udkm toolbox is currently not meant to provide any high-profile code here.
Optionally, we provide an example of the microscopic 3-temperature-model by \citet{Koopmans2010} in the documentation.
This model provides decent fits of the average magnetization amplitude for the early delays after photoexcitation but already neglects thermal transport by phonons.
Instead of implementing complex spin dynamics simulations here, we decided to provide a generic interface to inject results from external magnetization dynamic calculations \cite{elk, Atxitia2017, Ubermag, oommf, mumax3, spilady} or to easily add user-defined code.
For that we provide the \code{Magnetization} interface class.
By inheriting from this interface class the user can reuse all features of the \code{Simulation} class, such as caching, and only needs to overwrite the \code{calc_magnetization_map()} method.
This method can have the \code{temp_map} and \code{strain_map} as inputs and can access all properties of the sample \code{Structure}.
Its return value must be a \code{magnetization_map}, which contains the magnetization amplitude and direction ($\gamma$ and $\phi$ in radians), see Fig.~\ref{fig:experiment}, for every layer at every delay of the dynamic simulation.

\begin{figure}[tb]
    \centering
    \includegraphics[width=\figurewidth\textwidth]{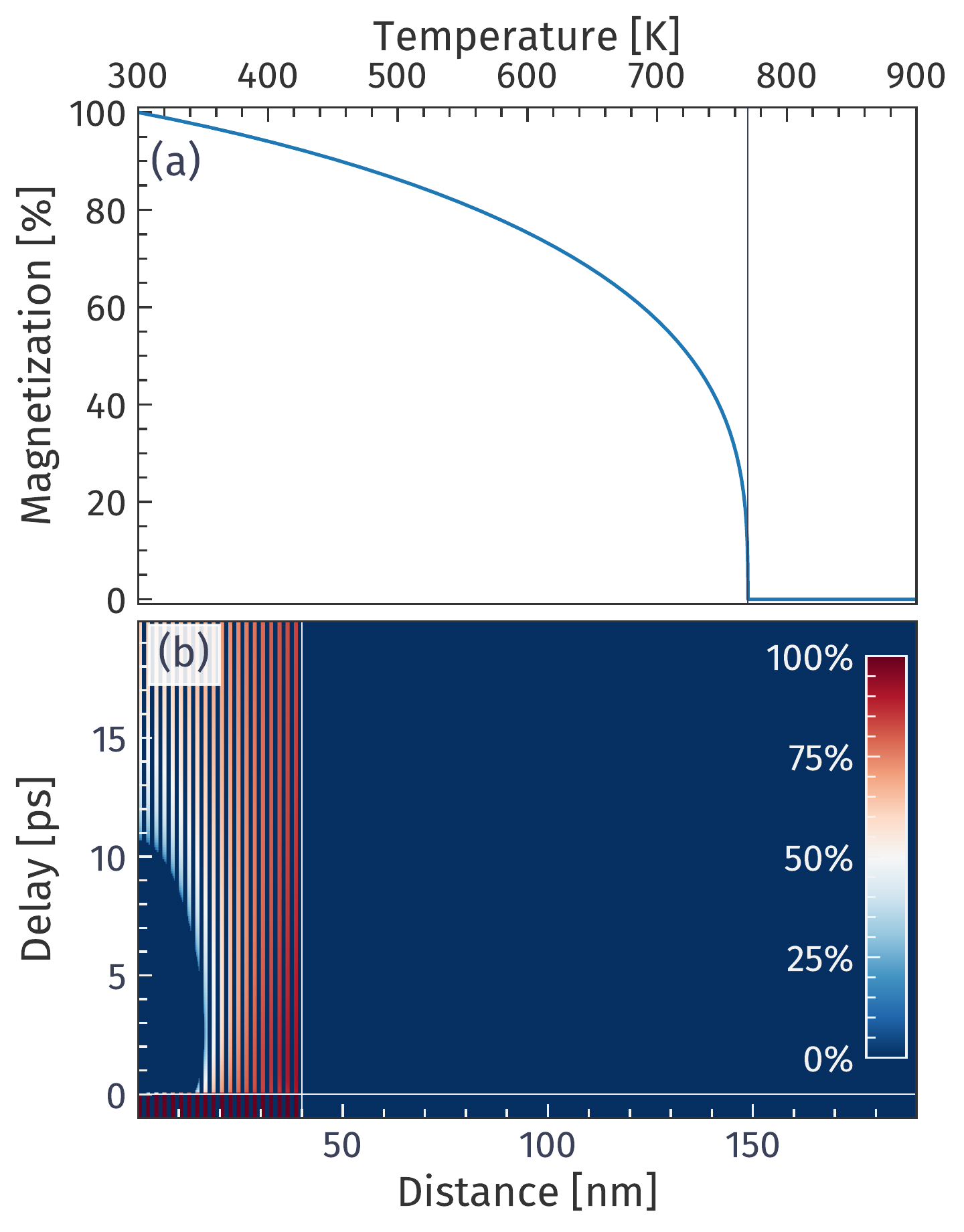}
    \caption{Spatio-temporal map of the magnetization amplitude.
    The magnetization amplitude of the Fe layers is completely quenched for high local temperatures above $T_C = 770$~K.
    }
    \label{fig:magnetization}
\end{figure}

As an example we employ the Bloch formula, c.f. Eq.~\ref{eq:bloch}, to calculate the spontaneous magnetization $M$ as function of the layer temperature $T$ for every delay.
\begin{equation}
    M(T) = M_0((1-(T/T_C)^{\alpha})^\beta)
    \label{eq:bloch}
\end{equation}
We use the parameters $T_C = 770$~K, $\alpha=3/2$ and $\beta=0.32$ and set $M_0$ such that $M(300K) = 1$.
Running the calculation is accomplished by:
\begin{minted}{python}
mag = Magnetization(S, True)
magnetization_map = mag.get_magnetization_map(
    delays, strain_map=strain_map, temp_map=temp_map)
\end{minted}
The amplitude of the resulting \code{magnetization_map} is shown in Fig.~\ref{fig:magnetization}.

\subsubsection{Dynamical Magnetic X-ray Scattering}

In the last step we are going to calculate the static resonant magnetic scattering response of the sample before we use the \code{strain_map} and \code{magnetization_map} from the calculations above to simulate a dynamical scattering response similar to a real pump-probe experiment.
The \code{xray} module of the \udkm toolbox currently includes three different X-ray scattering formalisms.
A kinematical and dynamical X-ray scattering formalism is implemented in the \code{XrayKin} and \code{XrayDyn} classes, respectively.
They work only for crystalline unit cells and do not include magnetic scattering effects.
Therefor we use here the dynamical magnetic formalism of the \code{XrayDynMag} class, which is one of the most important new features of the \udkm toolbox.
This formalism has been adapted form the open-source Dyna Project \cite{dyna, Elzo2012} and was further improved by adding parallelization features, increasing the vectorization, and allowing for simultaneous energy- and $q_z$-dependent calculation of the reflectivity and transmissivity for both amorphous layers and crystalline unit cells.
For the later case no Bragg peaks from mono-atomic unit cells can be calculated as the Dyna formalism relies on scattering differences between adjacent layers.

\begin{figure}[tb]
    \centering
    \includegraphics[width=\figurewidth\textwidth]{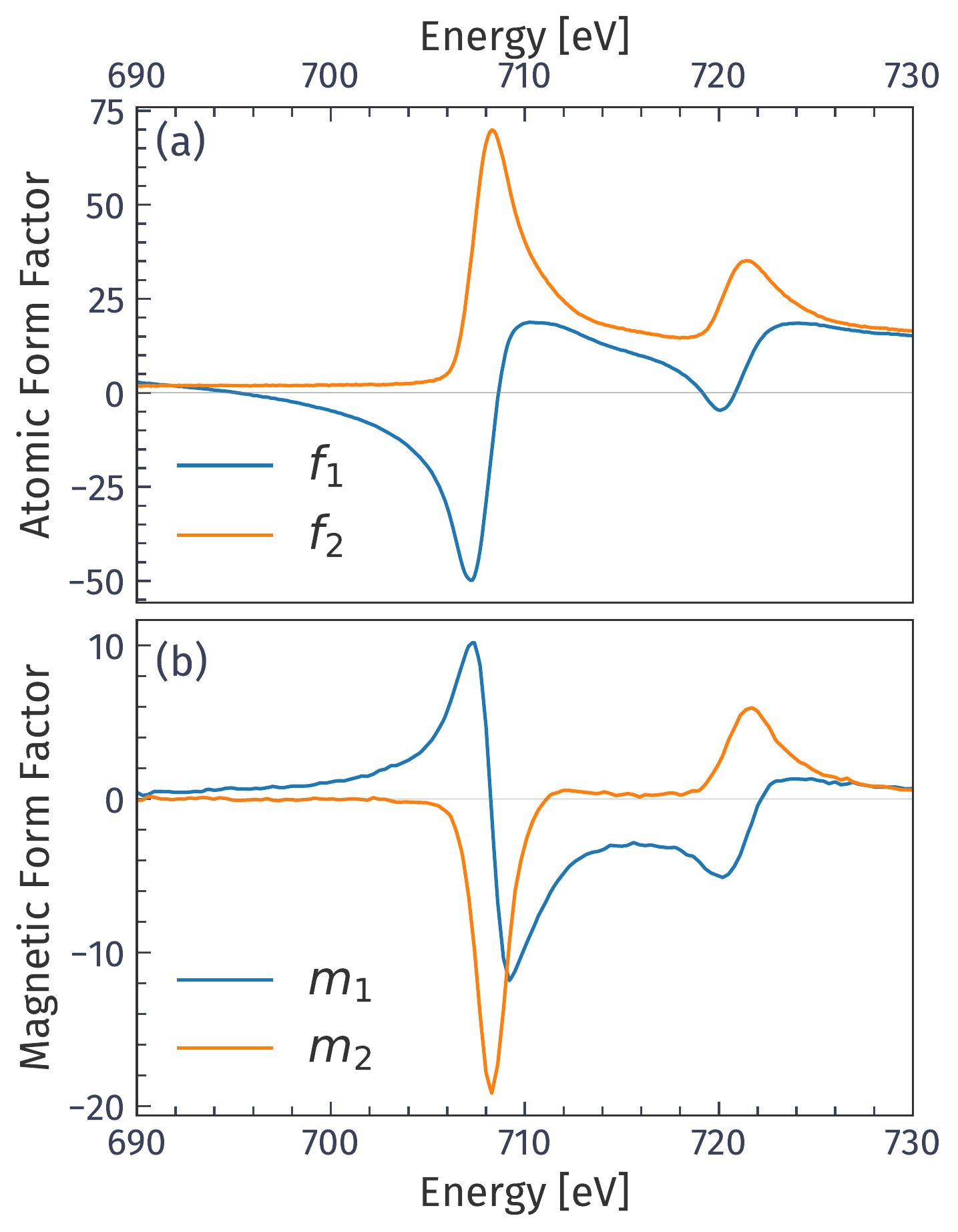}
    \caption{Energy-dependence of the refractive $f_1, m_1$ and absorptive $f_2, m_2$ parts of the atomic and magnetic form factors of Fe.
    }
    \label{fig:xrays_form_factors}
\end{figure}

In order to perform resonant magnetic scattering simulations the atomic and magnetic form factors need to be carefully determined.
This is usually done by measuring the dichroic absorption of circularly polarized light for the ferromagnetic state of the according element across the resonances of interest.
The average and difference spectra can then be carefully calibrated and scaled to theoretical values \cite{Chantler2001, henk1993a} in order to determine the absorptive parts $f_2$ and $m_2$.
By applying the Kramers-Kronig-relation the refractive contributions $f_1$ and $m_1$ can be retrieved as well.
This procedure is currently not within the scope of the \udkm toolbox but can be easily achieved with other free packages \cite{kkcalc, Watts:14}.
Figure~\ref{fig:xrays_form_factors} shows the atomic and magnetic form factors for the current simulations.

\begin{figure}[tb]
    \centering
    \includegraphics[width=\figurewidth\textwidth]{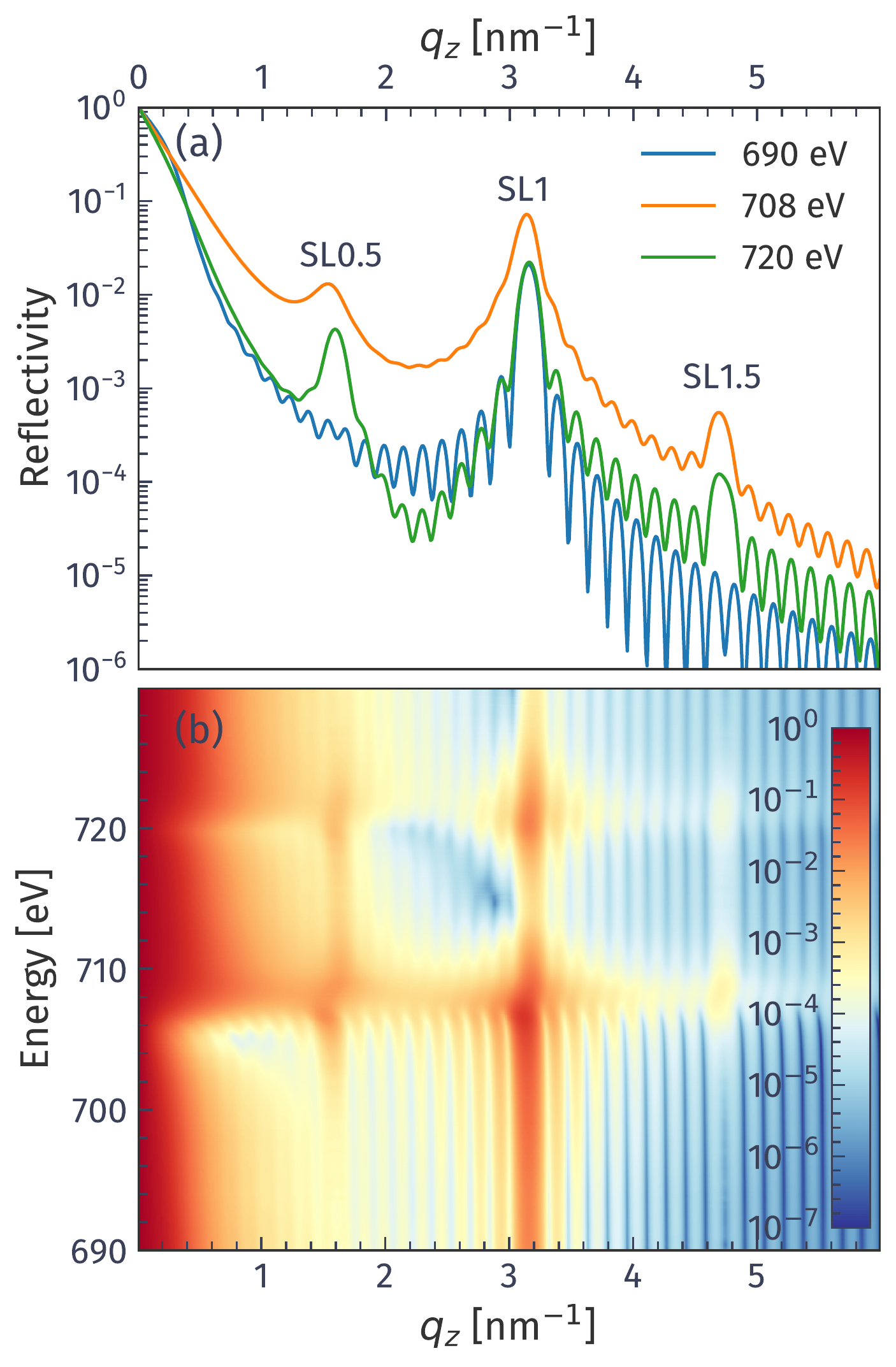}
    \caption{Energy and $q_z$-dependent scattering from the AFM-coupled Fe/Cr superlattice.
    Selected line-outs from panel (b) before as well as at the $L_3$ and $L_2$ resonances of Fe are shown in panel (a).
    The AFM Bragg peaks only appear on-resonance at $q_z^\text{SL0.5} = 1.56\ \text{nm}^{-1}$ and $q_z^\text{SL1.5} = 4.7\ \text{nm}^{-1}$, while the structural Bragg peak is present for all energies at $q_z^\text{SL1} = 3.13\ \text{nm}^{-1}$.
    }
    \label{fig:xrays_static}
\end{figure}

The reflected intensity \code{R} of the static sample for a given range of photon energies and scattering vectors $q_z$  can be easily calculated by the \code{homogeneous_reflectivity()} method of the \code{XrayDynMag} class:
\begin{minted}{python}
dyn_mag = ud.XrayDynMag(S, True)
dyn_mag.energy = numpy.r_[690:730:0.1]*u.eV
dyn_mag.qz = numpy.r_[0.01:5:0.01]/u.nm
R, _, _, _ = dyn_mag.homogeneous_reflectivity()
\end{minted}
The resulting scattering map is shown in Fig.~\ref{fig:xrays_static} and features a strong structural SL Bragg peak around $q_z^\text{SL1} = 3.13\ \text{nm}^{-1}$ for all photon energies.
The purely AFM Bragg peaks around $q_z^\text{SL0.5} = 1.56\ \text{nm}^{-1}$ and $q_z^\text{SL1.5} = 4.7\ \text{nm}^{-1}$ appear only at the Fe $L_3$ and $L_2$ resonances which provide magnetic contrast.

\begin{figure}[tb]
    \centering
    \includegraphics[width=\figurewidth\textwidth]{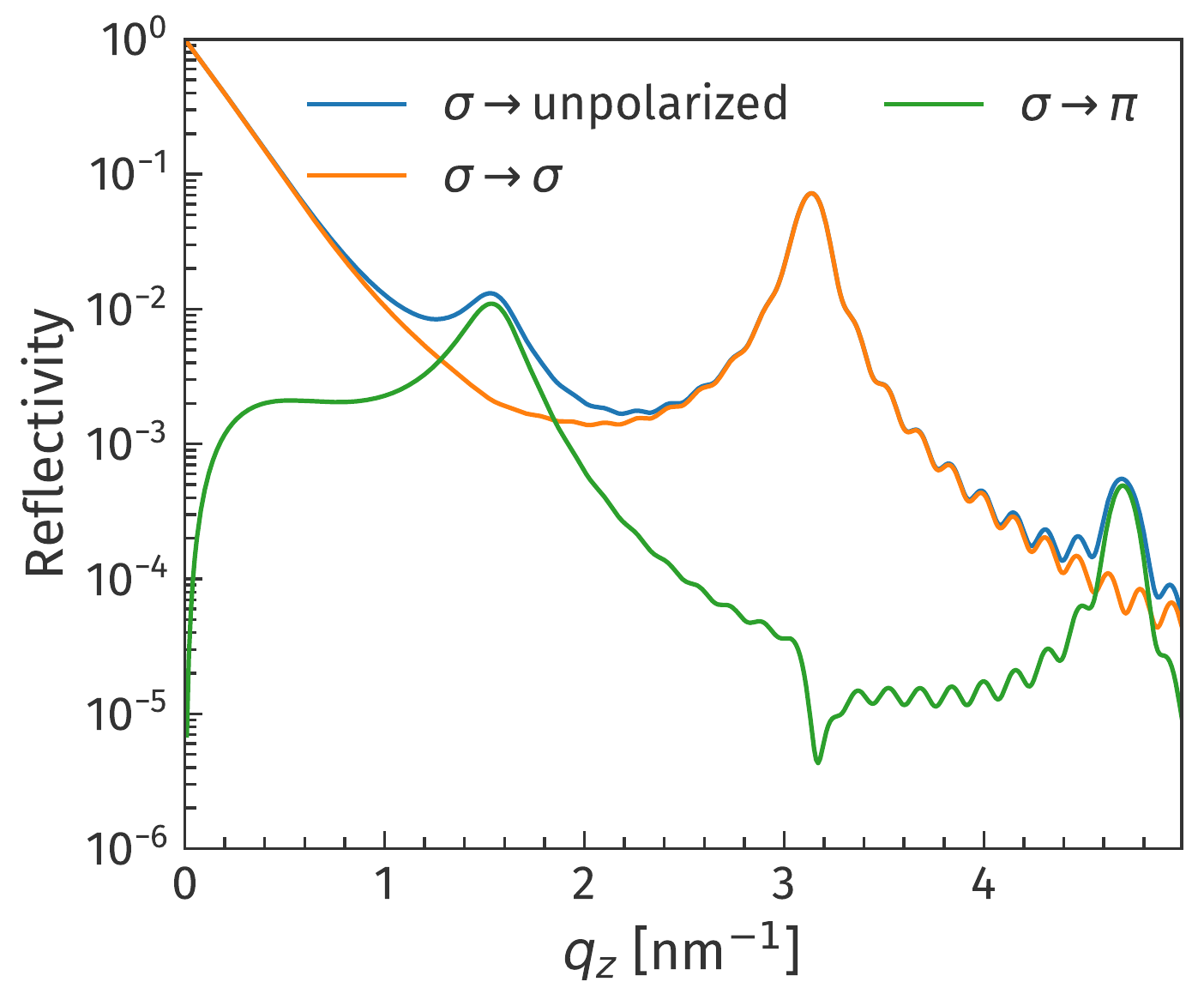}
    \caption{Polarization dependence of magnetic X-ray scattering from the Fe/Cr SL at 708~eV.
    The $\sigma \rightarrow \sigma$ channel only includes charge scattering and the $\sigma \rightarrow \pi$ channel only magnetic scattering.
    Without analyzer both contributions are included.
    }
    \label{fig:xrays_polarization}
\end{figure}

To verify the magnetic origin of the AFM Bragg peaks we conduct a polarization analysis within the dynamical magnetic X-ray scattering simulations.
Figure~\ref{fig:xrays_polarization} compares three different analyzer settings for fixed $\sigma$ polarization of the incident X-rays.
While normal charge scattering is not altering the polarization of the scattered light, this is not true for magnetic scattering.
Accordingly, without analyzer both charge and magnetic scattering are included as in Fig.~\ref{fig:xrays_static} before, the $\sigma \rightarrow \sigma$ channel only includes charge scattering and the $\sigma \rightarrow \pi$ channel only includes magnetic scattering.
As expected, the AFM Bragg peaks do not appear in the $\sigma \rightarrow \sigma$ channel.
Due to the symmetric layer thicknesses in the SL also even orders of the AFM Bragg peak are forbidden in the $\sigma \rightarrow \pi$ channel.

\begin{figure}[tb]
    \centering
    \includegraphics[width=\figurewidth\textwidth]{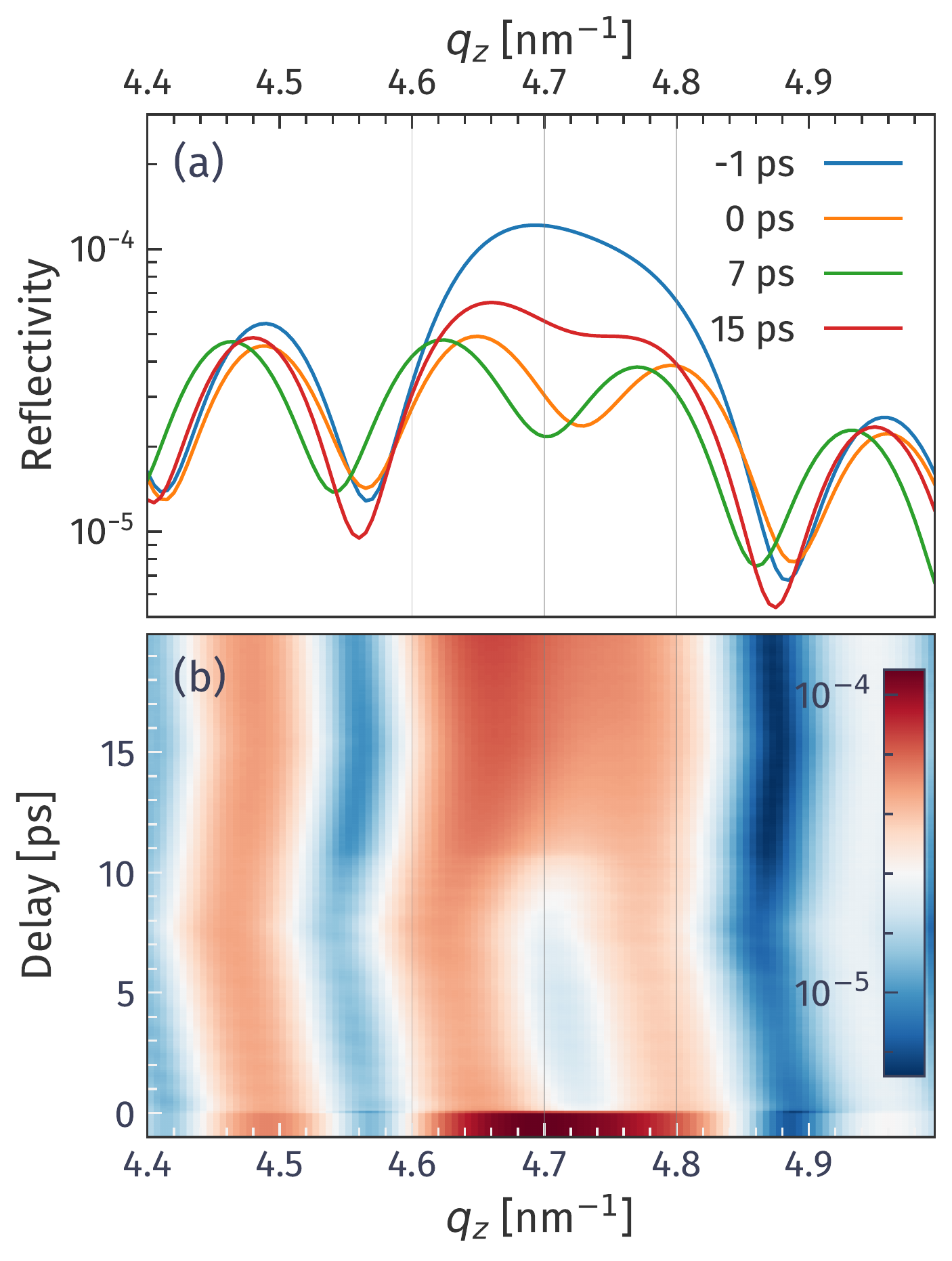}
    \caption{Transient resonant magnetic X-ray scattering around $q_z^\text{SL1.5} = 4.7\ \text{nm}^{-1}$ at 720~eV featuring a strong quenching and recovery of the AFM Bragg peak intensity as well as a peak shift due to coherent acoustic phonon dynamics.
    }
    \label{fig:xrays_dynamic}
\end{figure}

In the last step we calculate the dynamic response of the sample due to the above calculated spin and lattice dynamics, providing the \code{magnetization_map} and \code{strain_map} as inputs to the \code{inhomogeneous_reflectivity()} method of the \code{XrayDynMag} class:
\begin{minted}{python}
R_seq, _, _, _ = dyn_mag.inhomogeneous_reflectivity(
  strain_map = strain_map,
  magnetization_map=magnetization_map)
\end{minted}
The result around SL1.5 at 720~eV is shown in Fig.~\ref{fig:xrays_dynamic}.
Due the quasi-instantaneous demagnetization, as input by the \code{magnetization_map}, the AFM Bragg peak intensity is quenched and recovers on a 10~ps time scale.
At the same time coherent acoustic phonon dynamics lead to an obvious shift of the magnetic Bragg peak and the structural Laue oscillation beneath it.

\begin{figure}[tb]
    \centering
    \includegraphics[width=\figurewidth\textwidth]{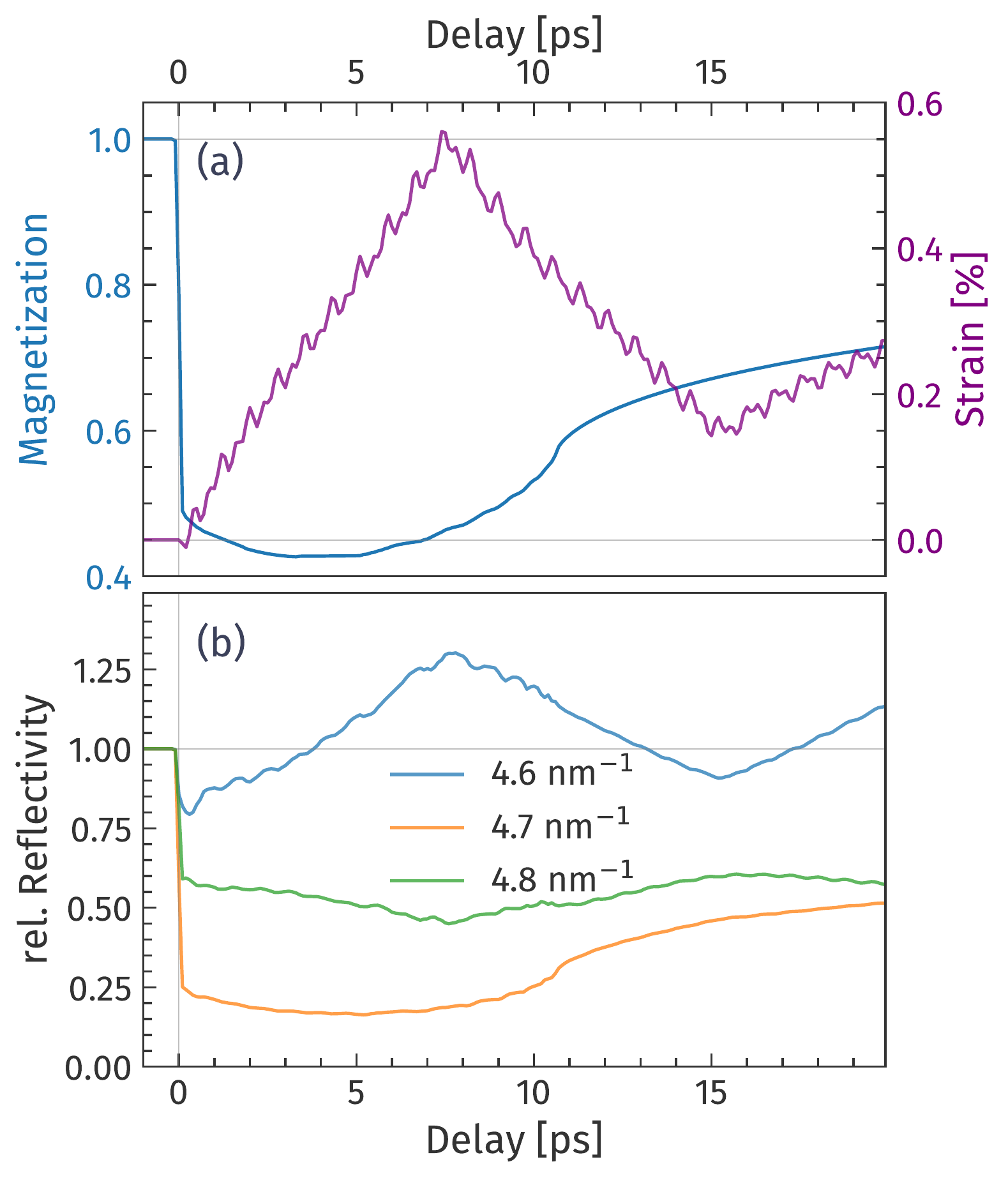}
    \caption{(a) Average magnetization of Fe layers and average strain of Fe and Cr layers within the SL structure.
    (b) Relative reflectivity at selected $q_z$ positions from Fig.~\ref{fig:xrays_dynamic}.
    }
    \label{fig:xrays_delay}
\end{figure}

In Fig.~\ref{fig:xrays_delay}~(b) we compare relative intensities $I(q_z, t)/I(q_z, t <0)$ for different scattering vectors $q_z$ with the input average strain and magnetization of within the Fe/Cr SL, c.f. panel~(a).
Depending on the selected $q_z$ the observed transients resemble dominantly the coherent acoustic phonon or magnetization dynamics or a mixture of both.
Note that even on the AFM Bragg peak maximum at $q_z = 4.7\ \text{nm}^{-1}$ the relative drop of the intensity by about 75~\% is not comparable with the change of the average magnetization in the Fe layers of less than 60~\%.
This results clearly indicates the importance of a careful analysis of time-resolved scattering data especially if magnetic effects are involved.

\section{Conclusions \& Outlook}

With the porting of the \udkm toolbox from \textsc{matlab} (MathWorks Inc.) to Python several new features have been added to the toolbox.
The implementation of magnetic properties as well as the resonant magnetic scattering formalism drastically extends its outreach.
The toolbox allows to carry out a full set of static and dynamics simulations on one and the same sample structure without switching between packages or even programming languages.
At the same time it relies to 100~\% on free and open-source software making it accessible to a broader audience in research and teaching.
This is further supported by its ease of use and detailed documentation.

The open-source aspect of the \udkm toolbox allows for easy adaption and adding extensions to it.
These can be either done by the users themselves or discussed at the project page at \href{https://github.com/dschick/udkm1Dsim}{\nolinkurl{github.com/dschick/udkm1Dsim}}.
Possible new features can be interfaces to other existing simulation packages, such as Ubermag \cite{Ubermag} in order to allow for more detailed magnetization dynamics.
Also the inclusion of temperatures in the X-ray scattering simulations will be interesting in terms of Debye-Waller-effects and transient changes of the scattering factors due to excitation of the band structure.
Additional probing techniques, such as visible light or electron scattering could be included in the future as well.

\section{Acknowledgment}

We would like to acknowledge the discussions with Marc Herzog, Stéphane Grenier, Lukas Alber, Valentino Scalera, Vivek Unikandanunni, and Stefano Bonetti.
We further thank Loïc Le Guyader for providing us with the source code for the optical multilayer formalism as well as Lisa-Marie Kern, Martin Borchert, and Martin Hennecke for extensive testing of the toolbox.

\bibliographystyle{elsarticle-num-names}
\bibliography{bibliography}

\begin{thebibliography}{42}
\expandafter\ifx\csname natexlab\endcsname\relax\def\natexlab#1{#1}\fi
\providecommand{\url}[1]{\texttt{#1}}
\providecommand{\href}[2]{#2}
\providecommand{\path}[1]{#1}
\providecommand{\DOIprefix}{doi:}
\providecommand{\ArXivprefix}{arXiv:}
\providecommand{\URLprefix}{URL: }
\providecommand{\Pubmedprefix}{pmid:}
\providecommand{\doi}[1]{\href{http://dx.doi.org/#1}{\path{#1}}}
\providecommand{\Pubmed}[1]{\href{pmid:#1}{\path{#1}}}
\providecommand{\bibinfo}[2]{#2}
\ifx\xfnm\relax \def\xfnm[#1]{\unskip,\space#1}\fi
\bibitem[{Waldecker et~al.(2016)Waldecker, Bertoni, Ernstorfer, and
  Vorberger}]{Waldecker2016}
\bibinfo{author}{L.~Waldecker}, \bibinfo{author}{R.~Bertoni},
  \bibinfo{author}{R.~Ernstorfer}, \bibinfo{author}{J.~Vorberger},
\newblock \bibinfo{title}{{Electron-phonon coupling and energy flow in a simple
  metal beyond the two-temperature approximation}},
\newblock \bibinfo{journal}{Phys. Rev. X} \bibinfo{volume}{6}
  (\bibinfo{year}{2016}) \bibinfo{pages}{1--11}.
  \DOIprefix\doi{10.1103/PhysRevX.6.021003}.
  \href{http://arxiv.org/abs/1507.03743}{{\tt arXiv:1507.03743}}.
\bibitem[{Schick et~al.(2014)Schick, Herzog, Wen, Chen, Adamo, Gaal, Schlom,
  Evans, Li, and Bargheer}]{schi2014a}
\bibinfo{author}{D.~Schick}, \bibinfo{author}{M.~Herzog},
  \bibinfo{author}{H.~Wen}, \bibinfo{author}{P.~Chen},
  \bibinfo{author}{C.~Adamo}, \bibinfo{author}{P.~Gaal}, \bibinfo{author}{D.~G.
  Schlom}, \bibinfo{author}{P.~G. Evans}, \bibinfo{author}{Y.~Li},
  \bibinfo{author}{M.~Bargheer},
\newblock \bibinfo{title}{{Localized Excited Charge Carriers Generate Ultrafast
  Inhomogeneous Strain in the Multiferroic BiFeO3}},
\newblock \bibinfo{journal}{Phys. Rev. Lett.} \bibinfo{volume}{112}
  (\bibinfo{year}{2014}) \bibinfo{pages}{097602}. \URLprefix
  \url{http://link.aps.org/doi/10.1103/PhysRevLett.112.097602}.
  \DOIprefix\doi{10.1103/PhysRevLett.112.097602}.
\bibitem[{Nicoul et~al.(2011)Nicoul, Shymanovich, Tarasevitch, von~der Linde,
  and Sokolowski-Tinten}]{nico2011a}
\bibinfo{author}{M.~Nicoul}, \bibinfo{author}{U.~Shymanovich},
  \bibinfo{author}{A.~Tarasevitch}, \bibinfo{author}{D.~von~der Linde},
  \bibinfo{author}{K.~Sokolowski-Tinten},
\newblock \bibinfo{title}{{Picosecond acoustic response of a laser-heated
  gold-film studied with time-resolved x-ray diffraction}},
\newblock \bibinfo{journal}{Appl. Phys. Lett.} \bibinfo{volume}{98}
  (\bibinfo{year}{2011}) \bibinfo{pages}{191902}. \URLprefix
  \url{http://link.aip.org/link/APPLAB/v98/i19/p191902/s1{\&}Agg=doi}.
  \DOIprefix\doi{10.1063/1.3584864}.
\bibitem[{Willems et~al.(2020)Willems, {von Korff Schmising}, Str{\"{u}}ber,
  Schick, Engel, Dewhurst, Elliot, Sharma, and Eisebitt}]{Willems2020}
\bibinfo{author}{F.~Willems}, \bibinfo{author}{C.~{von Korff Schmising}},
  \bibinfo{author}{C.~Str{\"{u}}ber}, \bibinfo{author}{D.~Schick},
  \bibinfo{author}{D.~Engel}, \bibinfo{author}{J.~Dewhurst},
  \bibinfo{author}{P.~Elliot}, \bibinfo{author}{S.~Sharma},
  \bibinfo{author}{E.~Eisebitt},
\newblock \bibinfo{title}{{Optical inter-site spin transfer probed by energy
  and spin-resolved transient absorption spectroscopy}},
\newblock \bibinfo{journal}{Nat. Commun.} \bibinfo{volume}{accepted}
  (\bibinfo{year}{2020}).
\bibitem[{You et~al.(2018)You, Tengdin, Chen, Shi, Zusin, Zhang, Gentry,
  Blonsky, Keller, Oppeneer, Kapteyn, Tao, and Murnane}]{You}
\bibinfo{author}{W.~You}, \bibinfo{author}{P.~Tengdin},
  \bibinfo{author}{C.~Chen}, \bibinfo{author}{X.~Shi},
  \bibinfo{author}{D.~Zusin}, \bibinfo{author}{Y.~Zhang},
  \bibinfo{author}{C.~Gentry}, \bibinfo{author}{A.~Blonsky},
  \bibinfo{author}{M.~Keller}, \bibinfo{author}{P.~M. Oppeneer},
  \bibinfo{author}{H.~Kapteyn}, \bibinfo{author}{Z.~Tao},
  \bibinfo{author}{M.~Murnane},
\newblock \bibinfo{title}{Revealing the nature of the ultrafast magnetic phase
  transition in ni by correlating extreme ultraviolet magneto-optic and
  photoemission spectroscopies},
\newblock \bibinfo{journal}{Phys. Rev. Lett.} \bibinfo{volume}{121}
  (\bibinfo{year}{2018}) \bibinfo{pages}{077204}. \URLprefix
  \url{https://link.aps.org/doi/10.1103/PhysRevLett.121.077204}.
  \DOIprefix\doi{10.1103/PhysRevLett.121.077204}.
\bibitem[{Stamm et~al.(2007)Stamm, Kachel, Pontius, Mitzner, Quast, Holldack,
  Khan, Lupulescu, Aziz, Wietstruk, D{\"{u}}rr, and Eberhardt}]{stam2007a}
\bibinfo{author}{C.~Stamm}, \bibinfo{author}{T.~Kachel},
  \bibinfo{author}{N.~Pontius}, \bibinfo{author}{R.~Mitzner},
  \bibinfo{author}{T.~Quast}, \bibinfo{author}{K.~Holldack},
  \bibinfo{author}{S.~Khan}, \bibinfo{author}{C.~Lupulescu},
  \bibinfo{author}{E.~F. Aziz}, \bibinfo{author}{M.~Wietstruk},
  \bibinfo{author}{H.~a. D{\"{u}}rr}, \bibinfo{author}{W.~Eberhardt},
\newblock \bibinfo{title}{{Femtosecond modification of electron localization
  and transfer of angular momentum in nickel.}},
\newblock \bibinfo{journal}{Nat. Mater.} \bibinfo{volume}{6}
  (\bibinfo{year}{2007}) \bibinfo{pages}{740--3}. \URLprefix
  \url{http://www.ncbi.nlm.nih.gov/pubmed/17721541}.
  \DOIprefix\doi{10.1038/nmat1985}.
\bibitem[{Jal et~al.(2017)Jal, L{\'{o}}pez-Flores, Pontius, Fert{\'{e}},
  Bergeard, Boeglin, Vodungbo, L{\"{u}}ning, and Jaouen}]{Jal2017}
\bibinfo{author}{E.~Jal}, \bibinfo{author}{V.~L{\'{o}}pez-Flores},
  \bibinfo{author}{N.~Pontius}, \bibinfo{author}{T.~Fert{\'{e}}},
  \bibinfo{author}{N.~Bergeard}, \bibinfo{author}{C.~Boeglin},
  \bibinfo{author}{B.~Vodungbo}, \bibinfo{author}{J.~L{\"{u}}ning},
  \bibinfo{author}{N.~Jaouen},
\newblock \bibinfo{title}{{Structural dynamics during laser-induced ultrafast
  demagnetization}},
\newblock \bibinfo{journal}{Phys. Rev. B} \bibinfo{volume}{95}
  (\bibinfo{year}{2017}) \bibinfo{pages}{184422}. \URLprefix
  \url{http://link.aps.org/doi/10.1103/PhysRevB.95.184422}.
  \DOIprefix\doi{10.1103/PhysRevB.95.184422}.
  \href{http://arxiv.org/abs/1701.01375}{{\tt arXiv:1701.01375}}.
\bibitem[{Henighan et~al.(2016)Henighan, Trigo, Bonetti, Granitzka, Higley,
  Chen, Jiang, Kukreja, Gray, Reid, Jal, Hoffmann, Kozina, Song, Chollet, Zhu,
  Xu, Jeong, Carva, Maldonado, Oppeneer, Samant, Parkin, Reis, and
  D{\"{u}}rr}]{Henighan2016}
\bibinfo{author}{T.~Henighan}, \bibinfo{author}{M.~Trigo},
  \bibinfo{author}{S.~Bonetti}, \bibinfo{author}{P.~Granitzka},
  \bibinfo{author}{D.~Higley}, \bibinfo{author}{Z.~Chen},
  \bibinfo{author}{M.~P. Jiang}, \bibinfo{author}{R.~Kukreja},
  \bibinfo{author}{A.~Gray}, \bibinfo{author}{A.~H. Reid},
  \bibinfo{author}{E.~Jal}, \bibinfo{author}{M.~C. Hoffmann},
  \bibinfo{author}{M.~Kozina}, \bibinfo{author}{S.~Song},
  \bibinfo{author}{M.~Chollet}, \bibinfo{author}{D.~Zhu},
  \bibinfo{author}{P.~F. Xu}, \bibinfo{author}{J.~Jeong},
  \bibinfo{author}{K.~Carva}, \bibinfo{author}{P.~Maldonado},
  \bibinfo{author}{P.~M. Oppeneer}, \bibinfo{author}{M.~G. Samant},
  \bibinfo{author}{S.~S.~P. Parkin}, \bibinfo{author}{D.~A. Reis},
  \bibinfo{author}{H.~A. D{\"{u}}rr},
\newblock \bibinfo{title}{{Generation mechanism of terahertz coherent acoustic
  phonons in Fe}},
\newblock \bibinfo{journal}{Phys. Rev. B} \bibinfo{volume}{93}
  (\bibinfo{year}{2016}) \bibinfo{pages}{220301}. \URLprefix
  \url{https://link.aps.org/doi/10.1103/PhysRevB.93.220301}.
  \DOIprefix\doi{10.1103/PhysRevB.93.220301}.
\bibitem[{von Reppert et~al.(2016)von Reppert, Pudell, Koc, Reinhardt,
  Leitenberger, Dumesnil, Zamponi, and Bargheer}]{Reppert2016}
\bibinfo{author}{A.~von Reppert}, \bibinfo{author}{J.~Pudell},
  \bibinfo{author}{A.~Koc}, \bibinfo{author}{M.~Reinhardt},
  \bibinfo{author}{W.~Leitenberger}, \bibinfo{author}{K.~Dumesnil},
  \bibinfo{author}{F.~Zamponi}, \bibinfo{author}{M.~Bargheer},
\newblock \bibinfo{title}{{Persistent nonequilibrium dynamics of the thermal
  energies in the spin and phonon systems of an antiferromagnet}},
\newblock \bibinfo{journal}{Struct. Dyn.} \bibinfo{volume}{3}
  (\bibinfo{year}{2016}) \bibinfo{pages}{054302}. \URLprefix
  \url{http://dx.doi.org/10.1063/1.4961253}. \DOIprefix\doi{10.1063/1.4961253}.
\bibitem[{Anisimov et~al.(1975)Anisimov, Kapeliovich, and
  Perel'man}]{anis1974a}
\bibinfo{author}{S.~I. Anisimov}, \bibinfo{author}{B.~L. Kapeliovich},
  \bibinfo{author}{T.~L. Perel'man},
\newblock \bibinfo{title}{{Electron emission from metal surfaces exposed to
  ultrashort laser pulses}},
\newblock \bibinfo{journal}{J. Exp. Theor. Phys.} \bibinfo{volume}{39}
  (\bibinfo{year}{1975}) \bibinfo{pages}{375--377}.
\bibitem[{Koopmans et~al.(2010)Koopmans, Malinowski, {Dalla Longa}, Steiauf,
  F{\"{a}}hnle, Roth, Cinchetti, and Aeschlimann}]{Koopmans2010}
\bibinfo{author}{B.~Koopmans}, \bibinfo{author}{G.~Malinowski},
  \bibinfo{author}{F.~{Dalla Longa}}, \bibinfo{author}{D.~Steiauf},
  \bibinfo{author}{M.~F{\"{a}}hnle}, \bibinfo{author}{T.~Roth},
  \bibinfo{author}{M.~Cinchetti}, \bibinfo{author}{M.~Aeschlimann},
\newblock \bibinfo{title}{{Explaining the paradoxical diversity of ultrafast
  laser-induced demagnetization.}},
\newblock \bibinfo{journal}{Nat. Mater.} \bibinfo{volume}{9}
  (\bibinfo{year}{2010}) \bibinfo{pages}{259--265}. \URLprefix
  \url{http://dx.doi.org/10.1038/nmat2593}. \DOIprefix\doi{10.1038/nmat2593}.
\bibitem[{Alber et~al.(2020)Alber, Scalera, Unikandanunni, Schick, and
  Bonetti}]{Alber2020}
\bibinfo{author}{L.~Alber}, \bibinfo{author}{V.~Scalera},
  \bibinfo{author}{V.~Unikandanunni}, \bibinfo{author}{D.~Schick},
  \bibinfo{author}{S.~Bonetti}, \bibinfo{title}{{NTMpy: An open source package
  for solving coupled parabolic differential equations in the framework of the
  three-temperature model}}, \bibinfo{year}{2020}.
\bibitem[{Thomsen et~al.(1986)Thomsen, Grahn, Maris, and Tauc}]{thom1986a}
\bibinfo{author}{C.~Thomsen}, \bibinfo{author}{H.~Grahn},
  \bibinfo{author}{H.~Maris}, \bibinfo{author}{J.~Tauc},
\newblock \bibinfo{title}{{Surface generation and detection of phonons by
  picosecond light pulses}},
\newblock \bibinfo{journal}{Phys. Rev. B} \bibinfo{volume}{34}
  (\bibinfo{year}{1986}) \bibinfo{pages}{4129--4138}. \URLprefix
  \url{http://link.aps.org/doi/10.1103/PhysRevB.34.4129}.
  \DOIprefix\doi{10.1103/PhysRevB.34.4129}.
\bibitem[{Herzog et~al.(2012)Herzog, Schick, Gaal, Shayduk, {Korff Schmising},
  and Bargheer}]{herz2012b}
\bibinfo{author}{M.~Herzog}, \bibinfo{author}{D.~Schick},
  \bibinfo{author}{P.~Gaal}, \bibinfo{author}{R.~Shayduk},
  \bibinfo{author}{C.~{Korff Schmising}}, \bibinfo{author}{M.~Bargheer},
\newblock \bibinfo{title}{{Analysis of ultrafast X-ray diffraction data in a
  linear-chain model of the lattice dynamics}},
\newblock \bibinfo{journal}{Appl. Phys. A} \bibinfo{volume}{106}
  (\bibinfo{year}{2012}) \bibinfo{pages}{489--499}. \URLprefix
  \url{http://www.springerlink.com/content/e8605244673r1121/}.
  \DOIprefix\doi{10.1007/s00339-011-6719-z}.
\bibitem[{spi(2021)}]{spilady}
\bibinfo{title}{{SPILADY - A Spin-Lattice Dynamics Simulation Program}},
  \bibinfo{howpublished}{https://ccfe.ukaea.uk/resources/spilady},
  \bibinfo{year}{2021}.
\bibitem[{Ma et~al.(2016)Ma, Dudarev, and Woo}]{MA2016350}
\bibinfo{author}{P.-W. Ma}, \bibinfo{author}{S.~Dudarev},
  \bibinfo{author}{C.~Woo},
\newblock \bibinfo{title}{Spilady: A parallel cpu and gpu code for
  spin–lattice magnetic molecular dynamics simulations},
\newblock \bibinfo{journal}{Computer Physics Communications}
  \bibinfo{volume}{207} (\bibinfo{year}{2016}) \bibinfo{pages}{350--361}.
  \URLprefix
  \url{https://www.sciencedirect.com/science/article/pii/S0010465516301412}.
  \DOIprefix\doi{https://doi.org/10.1016/j.cpc.2016.05.017}.
\bibitem[{Atxitia et~al.(2017)Atxitia, Hinzke, and Nowak}]{Atxitia2017}
\bibinfo{author}{U.~Atxitia}, \bibinfo{author}{D.~Hinzke},
  \bibinfo{author}{U.~Nowak},
\newblock \bibinfo{title}{{Fundamentals and applications of the
  Landau-Lifshitz-Bloch equation}},
\newblock \bibinfo{journal}{J. Phys. D. Appl. Phys.} \bibinfo{volume}{50}
  (\bibinfo{year}{2017}). \DOIprefix\doi{10.1088/1361-6463/50/3/033003}.
\bibitem[{Ube(2021)}]{Ubermag}
\bibinfo{title}{{Ubermag}}, \bibinfo{howpublished}{https://ubermag.github.io},
  \bibinfo{year}{2021}.
\bibitem[{mum(2021)}]{mumax3}
\bibinfo{title}{{mumax3 GPU-accelerated micromagnetism}},
  \bibinfo{howpublished}{https://mumax.github.io/}, \bibinfo{year}{2021}.
\bibitem[{oom(2021)}]{oommf}
\bibinfo{title}{{The Object Oriented MicroMagnetic Framework (OOMMF) project at
  ITL/NIST}}, \bibinfo{howpublished}{https://math.nist.gov/oommf/},
  \bibinfo{year}{2021}.
\bibitem[{Vansteenkiste et~al.(2014)Vansteenkiste, Leliaert, Dvornik, Helsen,
  Garcia-Sanchez, and {Van Waeyenberge}}]{Vansteenkiste2014}
\bibinfo{author}{A.~Vansteenkiste}, \bibinfo{author}{J.~Leliaert},
  \bibinfo{author}{M.~Dvornik}, \bibinfo{author}{M.~Helsen},
  \bibinfo{author}{F.~Garcia-Sanchez}, \bibinfo{author}{B.~{Van Waeyenberge}},
\newblock \bibinfo{title}{{The design and verification of MuMax3}},
\newblock \bibinfo{journal}{AIP Adv.} \bibinfo{volume}{4}
  (\bibinfo{year}{2014}). \URLprefix \url{http://dx.doi.org/10.1063/1.4899186}.
  \DOIprefix\doi{10.1063/1.4899186}. \href{http://arxiv.org/abs/1406.7635}{{\tt
  arXiv:1406.7635}}.
\bibitem[{Ohta and Ishida(1990)}]{Ohta1990}
\bibinfo{author}{K.~Ohta}, \bibinfo{author}{H.~Ishida},
\newblock \bibinfo{title}{{Matrix formalism for calculation of the light beam
  intensity in stratified multilayered films, and its use in the analysis of
  emission spectra}},
\newblock \bibinfo{journal}{Appl. Opt.} \bibinfo{volume}{29}
  (\bibinfo{year}{1990}) \bibinfo{pages}{2466}. \URLprefix
  \url{https://www.osapublishing.org/abstract.cfm?URI=ao-29-16-2466}.
  \DOIprefix\doi{10.1364/AO.29.002466}.
\bibitem[{{Le Guyader} et~al.(2013){Le Guyader}, Kleibert, Nolting, Joly,
  Derlet, Pisarev, Kirilyuk, Rasing, and Kimel}]{LeGuyader2013}
\bibinfo{author}{L.~{Le Guyader}}, \bibinfo{author}{A.~Kleibert},
  \bibinfo{author}{F.~Nolting}, \bibinfo{author}{L.~Joly},
  \bibinfo{author}{P.~M. Derlet}, \bibinfo{author}{R.~V. Pisarev},
  \bibinfo{author}{A.~Kirilyuk}, \bibinfo{author}{T.~Rasing},
  \bibinfo{author}{a.~V. Kimel},
\newblock \bibinfo{title}{{Dynamics of laser-induced spin reorientation in
  Co/SmFeO3 heterostructure}},
\newblock \bibinfo{journal}{Phys. Rev. B} \bibinfo{volume}{87}
  (\bibinfo{year}{2013}) \bibinfo{pages}{054437}. \URLprefix
  \url{http://link.aps.org/doi/10.1103/PhysRevB.87.054437}.
  \DOIprefix\doi{10.1103/PhysRevB.87.054437}.
\bibitem[{Kriegner et~al.(2013)Kriegner, Wintersberger, and
  Stangl}]{Kriegner2013}
\bibinfo{author}{D.~Kriegner}, \bibinfo{author}{E.~Wintersberger},
  \bibinfo{author}{J.~Stangl},
\newblock \bibinfo{title}{{Xrayutilities: A versatile tool for reciprocal space
  conversion of scattering data recorded with linear and area detectors}},
\newblock \bibinfo{journal}{J. Appl. Crystallogr.} \bibinfo{volume}{46}
  (\bibinfo{year}{2013}) \bibinfo{pages}{1162--1170}.
  \DOIprefix\doi{10.1107/S0021889813017214}.
  \href{http://arxiv.org/abs/arXiv:1304.1732v1}{{\tt arXiv:arXiv:1304.1732v1}}.
\bibitem[{Stepanov et~al.(1998)Stepanov, Kondrashkina, K{\"{o}}hler, Novikov,
  Materlik, and Durbin}]{step1998a}
\bibinfo{author}{S.~Stepanov}, \bibinfo{author}{E.~Kondrashkina},
  \bibinfo{author}{R.~K{\"{o}}hler}, \bibinfo{author}{D.~Novikov},
  \bibinfo{author}{G.~Materlik}, \bibinfo{author}{S.~Durbin},
\newblock \bibinfo{title}{{Dynamical x-ray diffraction of multilayers and
  superlattices: Recursion matrix extension to grazing angles}},
\newblock \bibinfo{journal}{Phys. Rev. B} \bibinfo{volume}{57}
  (\bibinfo{year}{1998}) \bibinfo{pages}{4829--4841}. \URLprefix
  \url{http://link.aps.org/doi/10.1103/PhysRevB.57.4829}.
  \DOIprefix\doi{10.1103/PhysRevB.57.4829}.
\bibitem[{Windt(1998)}]{Windt1998}
\bibinfo{author}{D.~L. Windt},
\newblock \bibinfo{title}{{IMD—Software for modeling the optical properties
  of multilayer films}},
\newblock \bibinfo{journal}{Comput. Phys.} \bibinfo{volume}{12}
  (\bibinfo{year}{1998}) \bibinfo{pages}{360}. \URLprefix
  \url{http://scitation.aip.org/content/aip/journal/cip/12/4/10.1063/1.168689}.
  \DOIprefix\doi{10.1063/1.168689}.
\bibitem[{Als-Nielsen and McMorrow(2011)}]{alsn2001a}
\bibinfo{author}{J.~Als-Nielsen}, \bibinfo{author}{D.~McMorrow},
  \bibinfo{title}{{Elements of Modern X-ray Physics}}, \bibinfo{publisher}{John
  Wiley {\&} Sons, Inc.}, \bibinfo{address}{Hoboken, NJ, USA},
  \bibinfo{year}{2011}. \URLprefix
  \url{http://doi.wiley.com/10.1002/9781119998365}.
  \DOIprefix\doi{10.1002/9781119998365}.
\bibitem[{Elzo et~al.(2012)Elzo, Jal, Bunau, Grenier, Joly, Ramos, Tolentino,
  Tonnerre, and Jaouen}]{Elzo2012}
\bibinfo{author}{M.~Elzo}, \bibinfo{author}{E.~Jal},
  \bibinfo{author}{O.~Bunau}, \bibinfo{author}{S.~Grenier},
  \bibinfo{author}{Y.~Joly}, \bibinfo{author}{A.~Y. Ramos},
  \bibinfo{author}{H.~C. Tolentino}, \bibinfo{author}{J.~M. Tonnerre},
  \bibinfo{author}{N.~Jaouen},
\newblock \bibinfo{title}{{X-ray resonant magnetic reflectivity of stratified
  magnetic structures: Eigenwave formalism and application to a W/Fe/W
  trilayer}},
\newblock \bibinfo{journal}{J. Magn. Magn. Mater.} \bibinfo{volume}{324}
  (\bibinfo{year}{2012}) \bibinfo{pages}{105--112}. \URLprefix
  \url{http://dx.doi.org/10.1016/j.jmmm.2011.07.019}.
  \DOIprefix\doi{10.1016/j.jmmm.2011.07.019}.
\bibitem[{Macke and Goering(2014)}]{Macke2014}
\bibinfo{author}{S.~Macke}, \bibinfo{author}{E.~Goering},
\newblock \bibinfo{title}{{Magnetic reflectometry of heterostructures}},
\newblock \bibinfo{journal}{J. Phys. Condens. Matter} \bibinfo{volume}{26}
  (\bibinfo{year}{2014}) \bibinfo{pages}{363201}. \URLprefix
  \url{http://www.ncbi.nlm.nih.gov/pubmed/25121937}.
  \DOIprefix\doi{10.1088/0953-8984/26/36/363201}.
\bibitem[{Schick et~al.(2014)Schick, Bojahr, Herzog, Shayduk, {von Korff
  Schmising}, and Bargheer}]{schi2013c}
\bibinfo{author}{D.~Schick}, \bibinfo{author}{A.~Bojahr},
  \bibinfo{author}{M.~Herzog}, \bibinfo{author}{R.~Shayduk},
  \bibinfo{author}{C.~{von Korff Schmising}}, \bibinfo{author}{M.~Bargheer},
\newblock \bibinfo{title}{{udkm1Dsim—A simulation toolkit for 1D ultrafast
  dynamics in condensed matter}},
\newblock \bibinfo{journal}{Comput. Phys. Commun.} \bibinfo{volume}{185}
  (\bibinfo{year}{2014}) \bibinfo{pages}{651--660}. \URLprefix
  \url{http://www.sciencedirect.com/science/article/pii/S0010465513003378}.
  \DOIprefix\doi{10.1016/j.cpc.2013.10.009}.
\bibitem[{udk(2021)}]{udkmML}
\bibinfo{title}{{udkm1Dsim Matlab}},
  \bibinfo{howpublished}{https://www.github.com/dschick/udkm1DsimML},
  \bibinfo{year}{2021}.
\bibitem[{Pin(2021)}]{Pint}
\bibinfo{title}{{Pint: makes units easy}},
  \bibinfo{howpublished}{https://pint.readthedocs.io/en/stable},
  \bibinfo{year}{2021}.
\bibitem[{Gr{\"{u}}nberg et~al.(1986)Gr{\"{u}}nberg, Schreiber, Pang, Brodsky,
  and Sowers}]{Grunberg1986}
\bibinfo{author}{P.~Gr{\"{u}}nberg}, \bibinfo{author}{R.~Schreiber},
  \bibinfo{author}{Y.~Pang}, \bibinfo{author}{M.~B. Brodsky},
  \bibinfo{author}{H.~Sowers},
\newblock \bibinfo{title}{{Layered Magnetic Structures: Evidence for
  Antiferromagnetic Coupling of Fe Layers across Cr Interlayers}},
\newblock \bibinfo{journal}{Phys. Rev. Lett.} \bibinfo{volume}{57}
  (\bibinfo{year}{1986}) \bibinfo{pages}{2442--2445}. \URLprefix
  \url{https://link.aps.org/doi/10.1103/PhysRevLett.57.2442}.
  \DOIprefix\doi{10.1103/PhysRevLett.57.2442}.
\bibitem[{Fullerton et~al.(1993)Fullerton, Conover, Mattson, Sowers, and
  Bader}]{Fullerton1993}
\bibinfo{author}{E.~E. Fullerton}, \bibinfo{author}{M.~J. Conover},
  \bibinfo{author}{J.~E. Mattson}, \bibinfo{author}{C.~H. Sowers},
  \bibinfo{author}{S.~D. Bader},
\newblock \bibinfo{title}{{Oscillatory interlayer coupling and giant
  magnetoresistance in epitaxial Fe/Cr(211) and (100) superlattices}},
\newblock \bibinfo{journal}{Phys. Rev. B} \bibinfo{volume}{48}
  (\bibinfo{year}{1993}) \bibinfo{pages}{15755--15763}.
  \DOIprefix\doi{10.1103/PhysRevB.48.15755}.
\bibitem[{Holy et~al.(1999)Holy, Pietsch, and Baumbach}]{holy1999a}
\bibinfo{author}{V.~Holy}, \bibinfo{author}{U.~Pietsch},
  \bibinfo{author}{T.~Baumbach}, \bibinfo{title}{{High-resolution X-ray
  scattering from thin films and multilayers}}, \bibinfo{publisher}{Springer
  Berlin / Heidelberg}, \bibinfo{year}{1999}.
\bibitem[{Nefedov et~al.(2005)Nefedov, Grabis, Bergmann, Radu, and
  Zabel}]{Nefedov2005}
\bibinfo{author}{A.~Nefedov}, \bibinfo{author}{J.~Grabis},
  \bibinfo{author}{A.~Bergmann}, \bibinfo{author}{F.~Radu},
  \bibinfo{author}{H.~Zabel},
\newblock \bibinfo{title}{{X-ray resonant magnetic scattering by Fe/Cr
  superlattices}},
\newblock \bibinfo{journal}{Superlattices Microstruct.} \bibinfo{volume}{37}
  (\bibinfo{year}{2005}) \bibinfo{pages}{99--106}.
  \DOIprefix\doi{10.1016/j.spmi.2004.07.004}.
\bibitem[{Chantler(2001)}]{Chantler2001}
\bibinfo{author}{C.~T. Chantler},
\newblock \bibinfo{title}{{Detailed tabulation of atomic form factors,
  photoelectric absorption and scattering cross section, and mass attenuation
  coefficients in the vicinity of absorption edges in the soft X-ray ( Z =
  30–36, Z = 60–89, E = 0.1–10 keV) – addressing convergence iss}},
\newblock \bibinfo{journal}{J. Synchrotron Radiat.} \bibinfo{volume}{8}
  (\bibinfo{year}{2001}) \bibinfo{pages}{1124--1124}. \URLprefix
  \url{http://scripts.iucr.org/cgi-bin/paper?S0909049501008305}.
  \DOIprefix\doi{10.1107/S0909049501008305}.
\bibitem[{Henke et~al.(1993)Henke, Gullikson, and Davis}]{henk1993a}
\bibinfo{author}{B.~Henke}, \bibinfo{author}{E.~Gullikson},
  \bibinfo{author}{J.~Davis},
\newblock \bibinfo{title}{{X-Ray Interactions: Photoabsorption, Scattering,
  Transmission, and Reflection at E = 50-30,000 eV, Z = 1-92}},
\newblock \bibinfo{journal}{At. Data Nucl. Data Tables} \bibinfo{volume}{54}
  (\bibinfo{year}{1993}) \bibinfo{pages}{181--342}. \URLprefix
  \url{http://linkinghub.elsevier.com/retrieve/pii/S0092640X83710132}.
  \DOIprefix\doi{10.1006/adnd.1993.1013}.
\bibitem[{dyn(2021)}]{dyna}
\bibinfo{title}{{Dyna Project}},
  \bibinfo{howpublished}{http://dyna.neel.cnrs.fr}, \bibinfo{year}{2021}.
\bibitem[{elk(2021)}]{elk}
\bibinfo{title}{{The Elk Code}},
  \bibinfo{howpublished}{http://elk.sourceforge.net}, \bibinfo{year}{2021}.
\bibitem[{kkc(2021)}]{kkcalc}
\bibinfo{title}{{KKcalc}},
  \bibinfo{howpublished}{https://github.com/benajamin/kkcalc},
  \bibinfo{year}{2021}.
\bibitem[{Watts(2014)}]{Watts:14}
\bibinfo{author}{B.~Watts},
\newblock \bibinfo{title}{Calculation of the kramers-kronig transform of x-ray
  spectra by a piecewise laurent polynomial method},
\newblock \bibinfo{journal}{Opt. Express} \bibinfo{volume}{22}
  (\bibinfo{year}{2014}) \bibinfo{pages}{23628--23639}. \URLprefix
  \url{http://www.opticsexpress.org/abstract.cfm?URI=oe-22-19-23628}.
  \DOIprefix\doi{10.1364/OE.22.023628}.

\end{thebibliography}

\end{document}